\documentclass[12pt,preprint]{aastex}
\def\p/{\mbox{$^1$}}
\def\pp/{\mbox{$^2$}}
\def\ppp/{\mbox{$^3$}}
\def\pppp/{\mbox{$^4$}}
\def\m/{\mbox{$^{-1}$}}
\def\mm/{\mbox{$^{-2}$}}
\def\mmm/{\mbox{$^{-3}$}}
\def\mmmm/{\mbox{$^{-4}$}}
\def\Ms/{\mbox{M$_\odot$}}

\def\bt{\mbox{$B_{\rm T}$}}
\def\zpbt{\mbox{ZP$_{B_{\rm T}}$}}
\def\vt{\mbox{$V_{\rm T}$}}
\def\zpvt{\mbox{ZP$_{V_{\rm T}}$}}
\def\btvt{\mbox{$B_{\rm T}-V_{\rm T}$}}
\def\zpbtvt{\mbox{ZP$_{B_{\rm T}-V_{\rm T}}$}}
\def\tj{\mbox{$V_{\rm T}-V$}}
\def\zptj{\mbox{ZP$_{V_{\rm T}-V}$}}
\def\by{\mbox{$b-y$}}
\def\zpby{\mbox{ZP$_{b-y}$}}
\def\m1{\mbox{$m_1$}}
\def\zpm1{\mbox{ZP$_{m_1}$}}
\def\c1{\mbox{$c_1$}}
\def\zpc1{\mbox{ZP$_{c_1}$}}
\def\zpv{\mbox{ZP$_{V}$}}
\def\bv{\mbox{$B-V$}}
\def\zpbv{\mbox{ZP$_{B-V}$}}
\def\ub{\mbox{$U-B$}}
\def\zpub{\mbox{ZP$_{U-B}$}}

\slugcomment{Accepted for publication in the Astronomical Journal}
\begin{document}

\title{A recalibration of optical photometry: \linebreak Tycho-2, Str\"omgren, and Johnson systems}
\shorttitle{A recalibration of optical photometry}

\author{J. Ma\'{\i}z Apell\'aniz\altaffilmark{1}}
\affil{Space Telescope Science Institute\altaffilmark{2}, 3700 San Martin 
Drive, Baltimore, MD 21218, U.S.A.}
\email{jmaiz@stsci.edu}


\altaffiltext{1}{Affiliated with the Space Telescope Division of the European 
Space Agency, ESTEC, Noordwijk, Netherlands.}
\altaffiltext{2}{The Space Telescope Science Institute is operated by the
Association of Universities for Research in Astronomy, Inc. under NASA
contract No. NAS5-26555.}

\begin{abstract}
I use high-quality HST spectrophotometry to analyze the calibration of three popular optical 
photometry systems: Tycho-2 \bt\vt, Str\"omgren $uvby$, and Johnson $UBV$. For Tycho-2, I revisit the
analysis of \citet{Maiz05b} to include the new recalibration of grating/aperture corrections, 
vignetting, and charge transfer inefficiency effects produced by the STIS group and to consider the 
consequences of both random and systematic uncertainties. The new results reaffirm the good
quality of both the Tycho-2 photometry and the HST spectrophotometry, but yield a slightly different
value for \zpbtvt\ of $0.033\pm 0.001$ (random) $\pm 0.005$ (systematic) magnitudes. For the Str\"omgren $v$, $b$,
and $y$ filters I find that the published sensitivity curves are consistent with the available photometry and
spectrophotometry and I derive new values for the associated \zpby\ and \zpm1. The same conclusion is drawn for the 
Johnson $B$ and $V$ filters and the associated \zpbv. The situation is different for the Str\"omgren $u$
and the Johnson $U$ filters. There I find that the published sensitivity curves yield results that are
inconsistent with the available photometry and spectrophotometry, likely caused by an incorrect 
treatment of atmospheric effects in the short-wavelength end. I reanalyze the data to 
produce new average sensitivity curves for those two filters and new values for \zpc1\ and \zpub. The new
computation of synthetic \ub\ and \bv\ colors uses a single $B$ sensitivity curve, which eliminates the previous
unphysical existence of different definitions for each color. Finally, I find that if one uses values from the literature
where uncertainties are not given, reasonable estimates for these are $1-2$\% for Str\"omgren \by, \m1, and \c1\ and
$2-3$\% for Johnson \bv\ and \ub. The use of the results in this article should lead to a 
significant reduction of systematic errors when comparing synthetic photometry models with real colors and indices.
\end{abstract}


\section{Introduction}

	The data explosion of the last decade has produced large quantities of photometric
measurements, either from dedicated ground-based projects (e.g. 2MASS, \citealt{Skruetal97}),
compilations from different sources
(e.g. GCPD, \citealt{Mermetal97}), or dedicated space-based missions (e.g.
Hipparcos, \citealt{ESA97}). The future promises an expansion of the phenomenon with the final
publication of data from projects such as the Sloan Digital Sky Survey \citep{Yorketal00} or GALEX 
\citep{Bianetal99} and new missions such as GAIA \citep{Mign05}. 

	The photometry from space-based missions is obtained with a fixed instrumental setup, does not 
suffer from perturbing atmospheric effects, and can be supported by a dedicated calibration program. 
On the other hand, observing from space also has its negative effects (e.g. the existence of Charge Transfer 
Inefficiency, or CTI, in CCD detectors due to radiation damage, \citealt{Kimbetal00}) but, overall, the available
data from recent space-based missions have very good photometric characteristics. The dedicated 
ground-based projects of the last decade also benefit from the uniformity of the instrumental setup, 
observing site, and reduction techniques and, though affected by the atmosphere, they can implement
extensive calibration programs and quality control mechanisms to ensure the stability of the obtained
data (see, e.g. \citealt{Coheetal03}, \citealt{Smitetal02}). Large dedicated photometric projects 
(either space- or ground-based) also benefit from the volume of the datasets and their coverage of a 
significant part of the sky (or all of it), which allow for statistical tests to detect possible 
systematic effects in the data.

	The situation is quite different for compilations of older ground-based data (see, e.g. 
\citealt{Lanz86,HaucMerm98}), which typically include tens or hundreds of studies, each one of them
with tens or hundreds of objects. The large number of sources implies different detectors, filters,
observatories, and reduction techniques, which undoubtedly introduce a scatter in the data and, quite
possibly, systematic errors (see \citealt{Stet05} and references therein for a detailed discussion on 
the problems associated with the reduction of ground-based photometry). Furthermore, in many cases the 
photometry obtained in the ``standard'' systems (e.g. Johnson or Str\"omgren) was established using equipment 
with lower precision than the one available today and using standard stars with a limited range of colors.
Therefore, it is possible that some systematic errors may have been introduced from the very
beginning, thus complicating a recalibration of the data. Those effects led \citet{Bessetal98} to
conclude that (a) standard systems may not represent any real linear system and (b) we should not be
reluctant to include ad-hoc corrections of a few percent to achieve an agreement between the observed and the
synthetic photometry in such systems. This rather pessimistic view may lead somebody to ask 
oneself why should we bother about the recalibration of photometric systems half-a-century old. The 
answer to that question is that many of the current calibrations are eventually tied up to those
standard systems, as evidenced by the fact that the most common way to simply characterize the
photometric properties of an object in a research proposal is to give its Johnson $V$ magnitude and \bv\
color. If we renounce to calibrate standard systems accurately we are condemned to have systematic
errors propagating down our reduction procedure.

	In this paper I take a more optimistic view of the problem by stating two points. In the first
place, even though it is true that some standard systems cannot be strictly characterized by a 
sensitivity (or throughput) curve and a zero point for each magnitude and/or color, 
it is also true that does not stop us from trying to derive the corresponding function and value that 
minimize the scatter of the data. In the second
place, some of the calibration problems encountered in the past may have been due not to
the photometry itself but to the data used to calibrate it, e.g. systematic errors in the measured 
spectrophotometry or in the assumed stellar model parameters. Therefore, the use of more modern 
calibration data may increase the accuracy. In other words, I believe it is possible to eliminate at 
least some of the systematic errors in the photometric calibration and to significantly lower the uncertainties in
the zero points.

	In paper I \citep{Maiz05b}, I used HST spectrophotometry to analyze Tycho-2 photometry in order
to check the accuracy of its published sensitivity curves and to calculate the zero points. In this
paper, I start by revisiting those data and then I use similar techniques to analyze the two most
common optical standard systems: Str\"omgren $uvby$ and Johnson $UBV$. In all cases, Vega will be used
as the reference spectrum to calculate magnitudes, colors, and indices (see Appendices).

\section{Spectrophotometry}

	A straightforward method to test the sensitivity curves and to calculate the zero points 
of filter systems is to observe a number of stars with well-characterized and sufficiently different 
spectral energy distributions (SEDs) and to compare the measured photometric magnitudes/colors/indices with those
derived from synthetic photometry\footnote{Alternatively, one can substitute the observed SEDs for synthetic 
(or model) ones, but doing so raises two issues: (a) the possible existence of errors in the SED modeling and/or in
the extinction characterization and (b) the need to obtain additional data to measure the relative flux 
calibration between the model SEDs and the real reference SED in Eq.~\ref{photon} in order to derive not only 
color zero points but also the zero point for a reference magnitude such as $V$. For those reasons, observed SEDs 
are in general preferred to model ones.}. If the sensitivity curves are correct, a plot of the difference
between the measured and synthetic values as a function of color or magnitude should yield a straight 
line with zero slope and an intercept equal to the zero point (see Appendices). If a slope is present
in such a plot, then the assumed sensitivity curves may be incorrect due to e.g. an inaccurate
characterization of atmospheric effects (which enter in the sensitivity curve for ground-based
observations in the reduction process to extrapolate to zero air masses) or an imprecise 
calibration of the detector. Of course, it is also possible that the cause of a non-zero slope lies in 
the SED library, and not in the photometry itself. Therefore, a necessary
preliminary step in the process is to ensure that the SEDs are as accurate as possible.

	The Space Telescope Imaging Spectrograph (STIS) onboard Hubble Space Telescope (HST) was able 
to provide during its seven years of operation the most accurate to-date spectrophotometry in the 
1\,150-10\,200 \AA\ range. The accuracy of its absolute flux calibration in the $V$ band is 
1.1\%, of which 0.7\% corresponds to the value of the absolute flux of Vega at the $V$ band 
\citep{Mege95} and 0.8\% to uncertainties\footnote{All of the uncertainties in this article are quoted 
as 1 $\sigma$.} in the photometry \citep{Bohl00,BohlGill04a}. On the other hand, the accuracy of
its relative flux calibration with respect to the $V$ band depends on the uncertainty of the
temperature of the three white dwarfs used as primary calibrators. This can be as high as 1-3\% in the
FUV but in the optical it is significantly lower due to the degeneracy of the spectral slopes at high temperatures. 
The accuracy of an individual STIS spectrophotometric observation is also limited 
by the instrument repeatability, which is 0.2-0.4\% for the wavelength range in which I
am interested in this paper \citep{Bohl00,BohlGill04b}.

	In this article I will use the same two STIS samples as in Paper I: the Next Generation
Spectral Library (NGSL; \citealt{Gregetal04}) and the Bohlin sample \citep{Bohletal01,BohlGill04b}. The 
reader is referred to Paper I for details. The NGSL sample is large ($\gtrsim 250$ non-variable
stars), is made up of relatively bright stars ($V$ between 2.5 and 11.5), and includes a large variety 
of SED types (sampling diverse temperatures, metallicities, and gravities), but was obtained with a 
STIS configuration (52X0.2 long slit at the E1 detector position and reading only a subarray of the 
CCD) which introduces some complications in the calibration. The Bohlin sample is smaller (19 stars),
consists of dimmer objects (only 2 are brighter than $V=9$, implying that only some of the stars have
reliable multicolor photometry available in the standard systems), 
and is made up mostly of early-type stars (only 5 stars 
have $\bv > 0.0$). The latter sample, however, uses the preferred STIS setup for absolute photometry 
(52X2 long slit at the center of the detector and reading the full CCD) and, in most cases, have 
repeated observations, thus allowing for a more straightforward calibration.

	Some of the calibration issues regarding the NGSL sample were discussed and dealt with in 
Paper I. Since that
article was published, the STIS group \citep{Prof05} has identified a number of issues that prompted me
to reprocess the data in order to improve the calibration. On the first place, it was discovered that
the effective throughput as a function of wavelength for a given long slit (e.g. 52X2 or 52X0.2) was not 
the same for all STIS gratings. This effect has been corrected in the STIS pipeline by introducing a 
grating/aperture correction table and making the appropriate changes in the {\tt calstis} software. 
Second, a new L-flat that incorporates the effects of vignetting and that should improve the flux 
calibration at the E1 detector position has been calculated. 
Finally, an error in the algorithm that calculates the CTI for subarrays has been fixed. All
of the above issues have no effect on the broadband colors of the Bohlin sample 
but can produce changes at the 1-2\% level for
the broadband colors of the NGSL sample. For that reason, before proceeding with an analysis of the
Str\"omgren and Johnson systems, I study the effects of the new calibration on the paper I results for
Tycho-2 photometry.

\section{Tycho-2 \bt\vt}

	As previously mentioned, the purpose of paper I was to test the published sensitivity curves
for the two Tycho-2 filters, \bt\ and \vt\ \citep{Bess00}, and to calculate their zero points. I found 
out in that paper that the sensitivity curves were indeed correct and derived a value of 
$\zpbt = 0.078 \pm 0.009$ from the Bohlin sample and of $\zpbtvt = 0.020 \pm 0.001$ from the NGSL
sample. The uncertainties quoted there are only the random ones and were derived by inverse variance
weighting from the photometry. The new calibration of the NGSL data can (and indeed does, as we will see next) 
affect the value for \zpbtvt\ but not that of \zpbt, which was derived from the Bohlin sample.

\begin{figure}[ht]
\centerline{\includegraphics*[width=\linewidth]{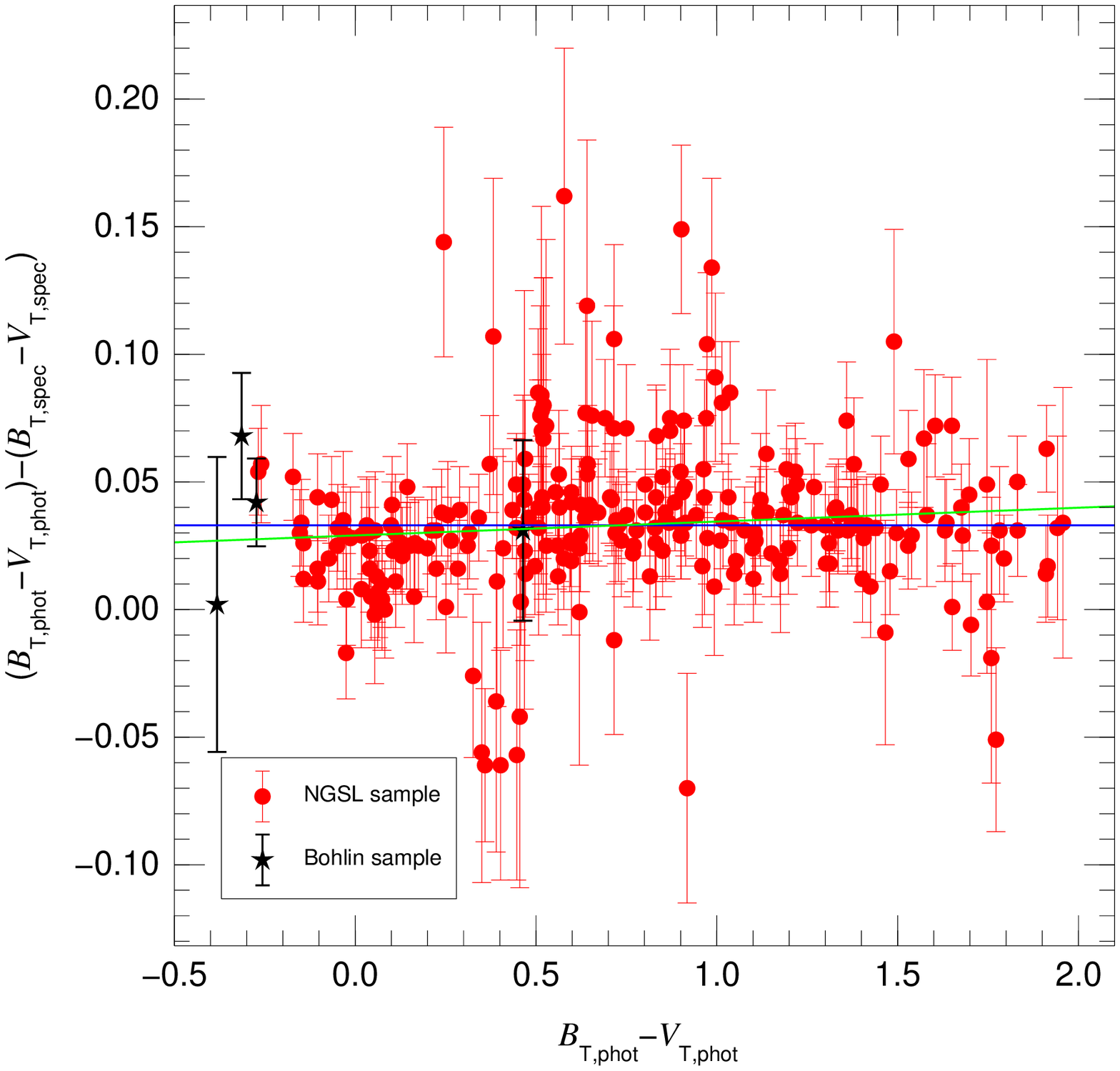}}
\caption{Comparison between photometric and spectrophotometric \btvt\ colors as a function 
of photometric \btvt\ for the NGSL + Bohlin samples. The error bars represent the photometric
uncertainties and the horizontal blue line marks the proposed \zpbtvt. The green line shows the result
of a weighted linear fit to the data.}
\label{btvtplot1}
\end{figure}

	I have reprocessed the NGSL data through {\tt calstis} and retained only the non-variable stars
with a 5\,600 \AA\ jump (the point where the G430L and G750L gratings overlap) of less than 3\% and
with uncertainties in the measured \bt\ and \vt\ magnitudes smaller than 0.06 mag. This
left 255 objects (as opposed to 256 in Paper I) for which the synthetic (or spectrophotometric) 
\bt\ and \vt\ magnitudes were 
computed. I show the difference between the photometric values (from the Tycho-2 catalog) and
spectrophotometric values (from the STIS spectra) of \btvt\ in Fig.~\ref{btvtplot1}, which is the
equivalent to Fig.~2 in Paper I\footnote{The figures in this paper that plot a photometric quantity against the 
difference between a photometric and a spectrophotometric value assume an associated zero point of 0.0 in 
Eq.~\ref{photon} because their purpose is to calculate the real zero point by analyzing the plots themselves.}.

\begin{figure}
\centerline{\includegraphics*[width=0.47\linewidth]{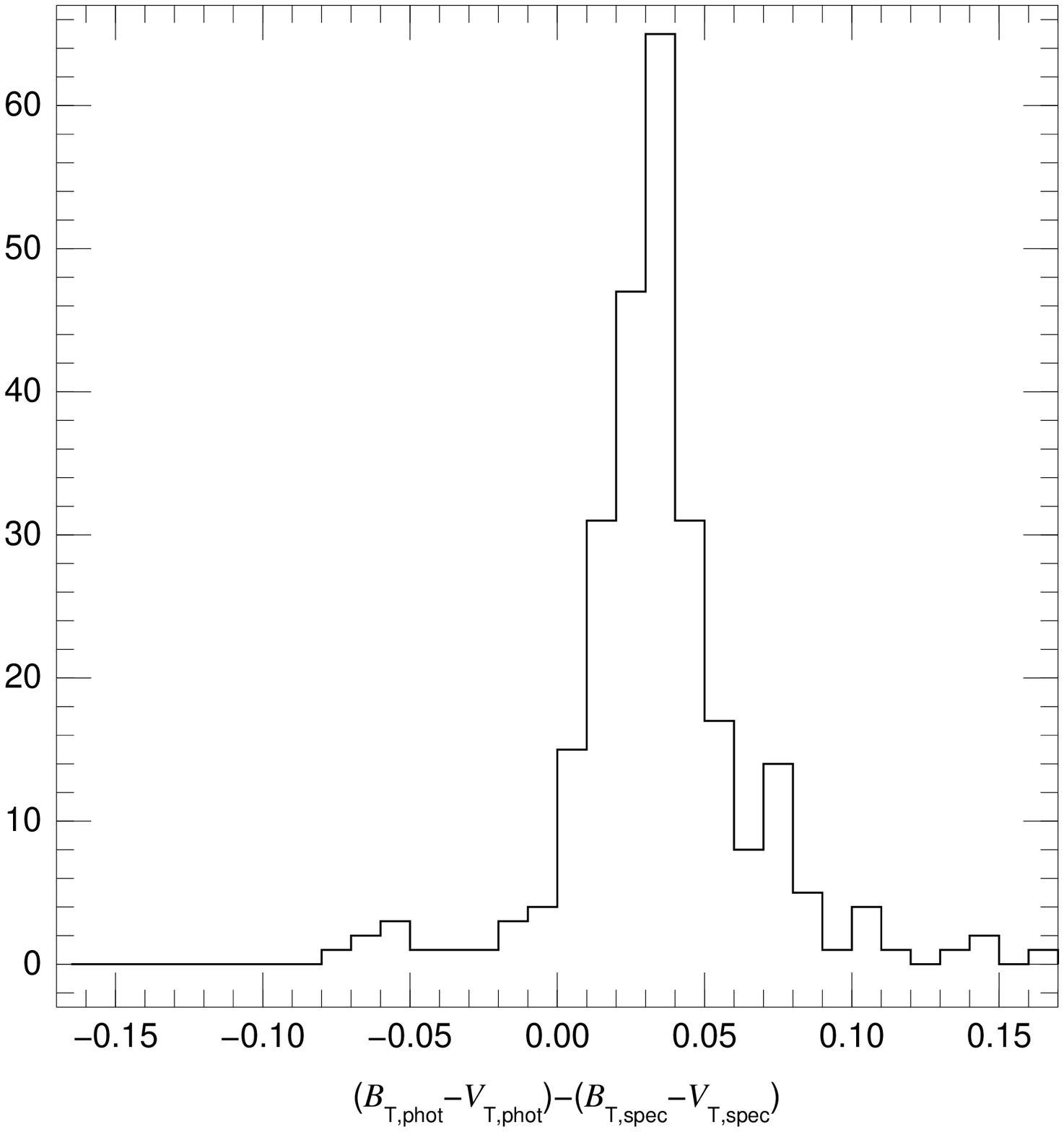}
            \includegraphics*[width=0.47\linewidth]{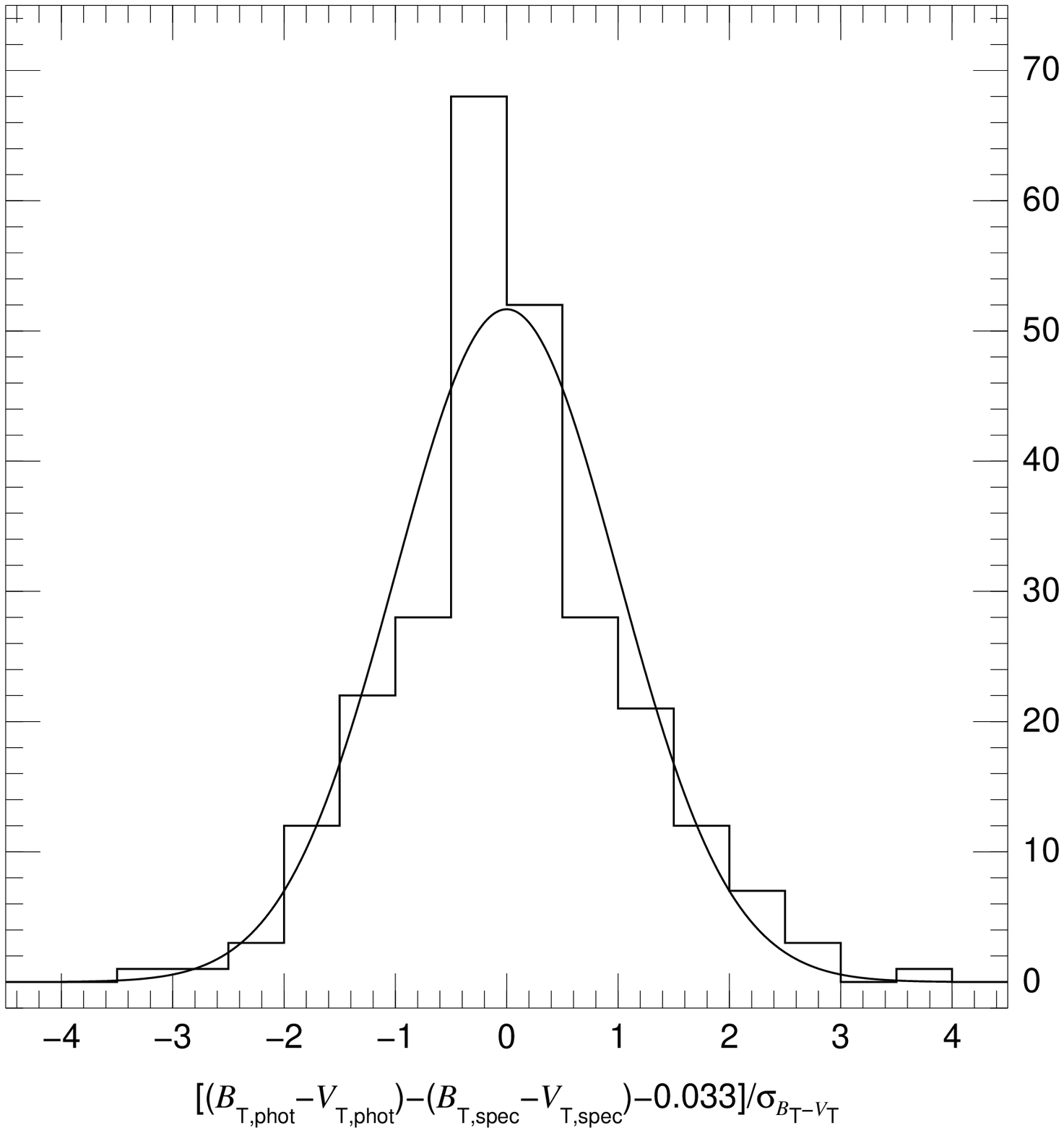}}
\caption{Histograms for the comparison between photometric and spectrophotometric \btvt\ colors
for the NGSL sample. (left) Regular histogram. (right) Histogram for the data shifted by the 
proposed \zpbtvt\ and normalized by the individual uncertainties. A Gaussian distribution
with $\mu=0$ and $\sigma=1$ is overplotted for comparison.}
\label{btvtplot2}
\end{figure}

	The main conclusion remains the same as in the previous paper: there is a very good
correspondence between the photometric and spectrophotometric \btvt\ colors, 
as evidenced by the nearly-flat
distribution of points in Fig.~\ref{btvtplot1}. The normalized histogram in Fig.~\ref{btvtplot2} also
corroborates that assertion: the residuals appear to be symmetrically distributed around the zero point
and the standard deviation of the histogram is 1.04, as in Paper I. Here I have also performed an
additional test by fitting a weighted straight line to the data in Fig.~\ref{btvtplot2}, which is shown
in green. The slope of the fit is nearly flat but not exactly so: $0.005\pm 0.002$. This implies that
either the Tycho-2 sensitivities or the calibration of the NGSL data are not perfect. However, the
effect is very small, since the deviations at the edges of the used color spectrum (which correspond
to low-reddening O and M stars, respectively) amount to only $\pm 0.006$ magnitudes, which is 1/3 of the
best individual photometric uncertainty for the colors in our sample. Also, $0.002/0.005 = 2.5$, i.e.
the slope is only 2.5 sigmas away from zero. Therefore, I do not consider 
necessary to derive new sensitivity curves for the Tycho-2 filters. Instead, this effect will be
included in the error analysis below as a systematic uncertainty\footnote{I consider an uncertainty in the zero
point to be systematic if it arises from errors in our knowledge of the characteristics of our photometric or 
spectrophotometric setup or reduction procedure, e.g. an incorrect mean sensitivity curve or offsets in the flux 
calibration of the spectrophotometry. I consider an uncertainty to be random if it would still be present in the
absence of any systematic effects, e.g. effects caused by a finite S/N of the photometry or spectrophotometry or 
random variations from the mean of the sensitivity curve due to changing atmospheric conditions between different 
observations.}.

	The new value I obtain for the \btvt\ zero point using inverse variance weighting is 
$\zpbtvt = 0.033\pm 0.001$, which is significantly larger than the value derived in Paper I 
($0.020\pm 0.001$). The 1.3\% difference between the two is within the expected range of the changes 
introduced in the new STIS calibration \citep{Prof05}. It is interesting to note that if I only use
the four Bohlin stars in Fig.~\ref{btvtplot1} to calculate \zpbtvt, I obtain a value of 
$0.046\pm 0.013$, which is at a distance of 1 sigma from the value proposed here but 2 sigmas away from
the previous one. Those values are consistent with an improvement in the new STIS calibration. The
uncertainty of 0.001 is a purely random one but, as we have seen before, there are also systematic
effects present. The slope to the fit described in the previous paragraph translates into a 1-sigma
uncertainty of 0.003 magnitudes. Furthermore, the uncertainty of 3\,000 K in the temperature of the white
dwarf calibrators quoted by \citet{BohlGill04b} translates into an additional (systematic) uncertainty of 
0.004 magnitudes, which was calculated using synthetic photometry of Kurucz models. Therefore, the proposed final 
value is $\zpbtvt = 0.033\pm 0.001$ (random) $\pm 0.005$ (systematic) magnitudes.

	As for the absolute calibration of Tycho-2 photometry, I defer its analysis until the Johnson $UBV$
photometry is analyzed.

\section{Str\"omgren $uvby$}

\subsection{Description and sample selection}

	The Str\"omgren standard system \citep{Mats69} has been the most commonly used 
intermediate-band photometric optical system in the last 40 years. It consists of 4 filters in the 
3\,000-6\,000 \AA\ range, $u$, $v$, $b$, and $y$ and the quantities that are presented in most works 
are:

\begin{eqnarray}
\by &   &            \\
\m1 & = & v - 2b + y \\
\c1 & = & u - 2v + b
\end{eqnarray}

	\by\ is a color similar (but not identical) 
to Johnson \bv\ or to Tycho-2 \btvt, \m1\ is an index that is most
sensitive to the metallicity of a star, and \c1\ is an index that measures the strength of the Balmer
jump \citep{Stro66}. \citet{HaucMerm98} have published a catalog of photometric measurements in the Str\"omgren 
system compiled from photoelectric data in the literature. In its current (summer 2005) version, available from the 
General Catalogue of Photometric Data (GCPD\footnote{\tt http://obswww.unige.ch/gcpd/gcpd.html}), it 
contains 66\,135 stars from 572 different sources. 

	I want to test whether the \citet{Mats69} sensitivity curves are consistent with the photometry
in the literature (in the sense of not having strong color terms when compared with the STIS spectrophotometry)
and to calculate the corresponding zero points for \by, \m1, and \c1. Note that the Str\"omgren GCPD data includes only 
one color and two indices, but not magnitudes (though Johnson $V$ is usually also given in many cases), so, as 
opposed to the Tycho-2 or Johnson cases, I only have to deal here with color/index zero points. We 
also need to consider that, as opposed to the Tycho-2 case, some of the scatter in the published photometric data 
is unavoidable due to the diversity of the sources. Therefore, our aims should be to obtain mean color/index values
for the stars in our samples that are representative of that diversity and to measure the intrinsic scatter. On
the other hand, I should not include in the sample highly discrepant values that are caused by
incorrect reductions or misidentifications because that would misrepresent the scatter that is really
unavoidable and may introduce biases in the interpretation of correct data. The elaboration of a 
photometric sample that is appropriate for a comparison with STIS spectrophotometry must allow both of
those principles to be followed by selecting stars for which there are enough data to discriminate
which values to include. With those ideas in mind, I followed these steps to create the Str\"omgren photometric
sample:

\begin{enumerate}
  \item I started by searching in the GCPD for the original Str\"omgren photometry (i.e. not the
	mean values) for all non-variable stars in the NGSL sample and for all stars in the Bohlin
	sample. As I did for the Tycho-2 case, for the NGSL sample only those stars with a 5\,600 \AA\ 
	jump in the spectrophotometry of less than 3\% were included.
  \item I calculated a weighted mean and a dispersion for \by, \m1, and \c1\ for each star, which are 
	taken to be the first estimate for the value and random uncertainty, respectively, for each
        of those color/indices.
  \item Highly discrepant individual data points (those more than 3 sigmas from the mean) were excluded 
	and the mean and dispersion recalculated for \by, \m1, and \c1\ for each star. 
  \item Only stars that had (a) at least three data points for each of \by, \m1, and \c1\ after the
	previous step and (b) random uncertainties of less than 0.035 magnitudes were kept in our 
	sample and the rest were excluded. This step provides a more
	meaningful value to the statistical weight that will be derived from the random uncertainty
	(i.e. it eliminates values that are likely to be biased or of lower quality).
  \item For the stars with Tycho-2 photometry, I plotted \by\ vs. \btvt\ (which should follow a quite
        tight correlation) and discarded the discrepant cases.
  \item For each of the three indices, the remaining stars were divided into two groups: (a) those with
        6 or more data points and (b) those with 3 to 5 data points. For the first group, those
	objects that had an uncertainty less than a critical value $\varepsilon_X$ (where $X$ = \by, 
	\m1, or \c1) had that uncertainty replaced by $\varepsilon_X$. For the second group, the
	same procedure was repeated but using $2\varepsilon_X$ as the replacement in order to account for the lower
	statistical weight that should be assigned to stars with a smaller number of data points.  
	There are two reasons for doing this. First, I avoid giving excessive statistical weights to 
	those cases where due to roundoff or special circumstances (e.g. most data points coming from 
	an identical instrumental setup and observing conditions) the dispersion of the data points is 
	too low. Second, I can vary $\varepsilon_X$ iteratively in order to get an accurate 
	measurement of the true intrinsic scatter by demanding that, once a zero point has been 
	calculated, the difference between the photometric and spectrophotometric values has a 
	reduced $\chi^2$ of 1.0 (or, alternatively, since the number of stars in our sample is large, that the 
	normalized histogram $(X_{\rm phot}-X_{\rm spec})/\sigma_X$ has a mean of zero and a 
	standard deviation of one). The initial values of $\varepsilon_{b-y}$, $\varepsilon_{m_1}$, and 
	$\varepsilon_{c_1}$ were obtained from \citet{HaucMerm98}\footnote{Note that their 
	definition of $\varepsilon_X$ is slightly different from the one here. This should not be an
	issue because I only use those values as initial guesses.}.
\end{enumerate}

	After the selection process was completed, I ended up with a total of 104 stars, 100 from the
NGSL sample and 4 from the Bohlin sample. In the next subsections I describe the results for \by, \m1, 
and \c1, respectively.

\subsection{\by}

\begin{figure}
\centerline{\includegraphics*[width=\linewidth]{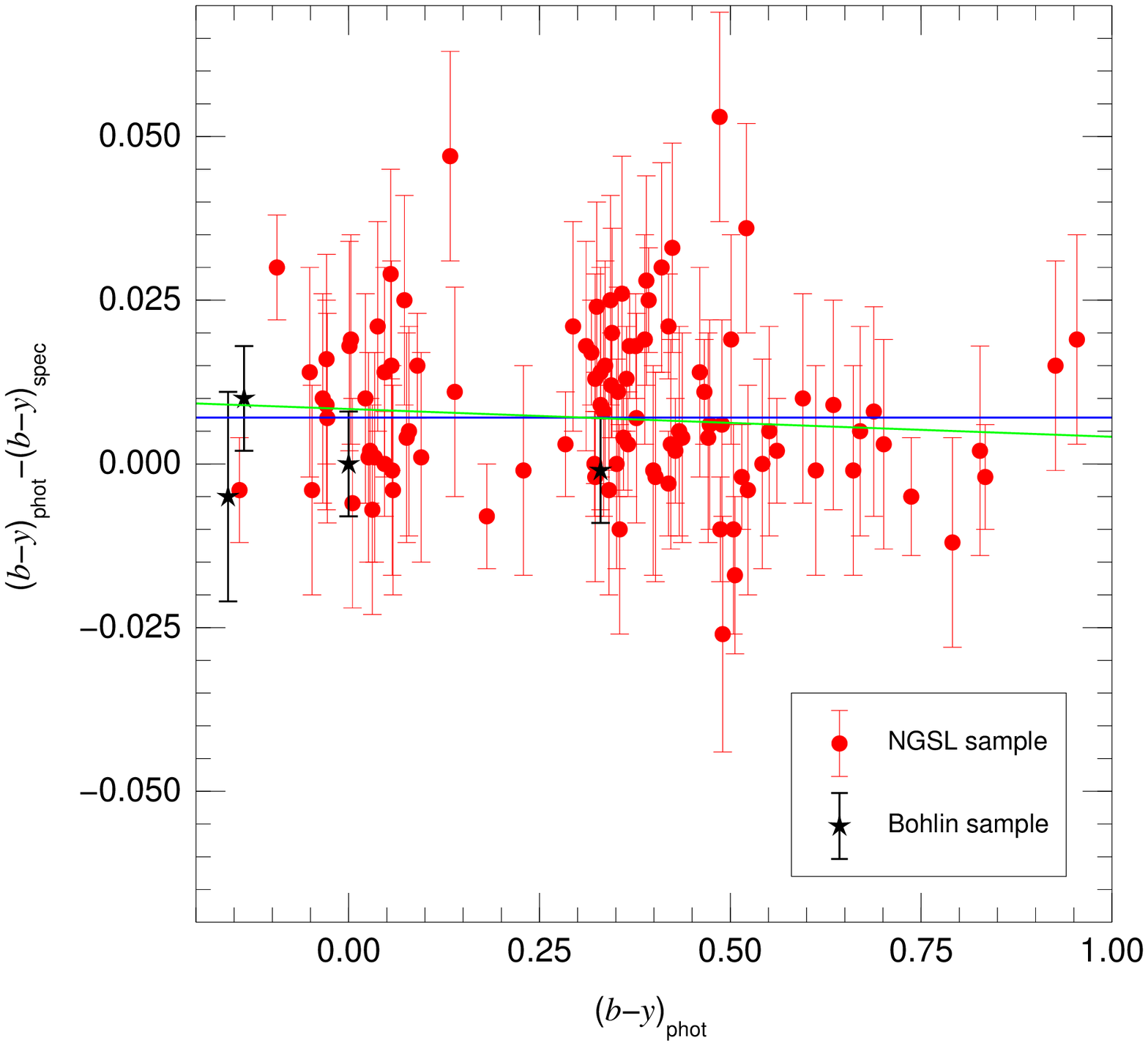}}
\caption{Comparison between photometric and spectrophotometric \by\ colors as a function 
of photometric \by\ for the NGSL + Bohlin samples. The error bars represent the photometric
uncertainties and the horizontal blue line marks the proposed \zpby. The green line shows the result
of a weighted linear fit to the data.}
\label{byplot}
\end{figure}

	The results for \by\ can be seen in Fig.~\ref{byplot}, where I have represented the difference
between the photometric and spectrophotometric values as a function of the photometric ones, in a
manner analogous to that of the previous section for \btvt. $\varepsilon_{b-y}$ was iteratively
determined to be 0.008 mag using the criterion stated in the previous subsection. 
The data does not manifest any significant trend with color and this shows in
the slope of the weighted linear fit, which is $-0.004\pm 0.005$ mag (less than 1 sigma away from
zero). Also, the slope of the fit has the opposite sign to the one I found for \btvt, suggesting
that the (small) discrepancies lie in the photometry, not in the STIS spectra. Furthermore, no obvious
different trend is seen between the objects in the NGSL and Bohlin samples. Therefore, I can 
conclude that the \citet{Mats69} sensitivity curves for $b$ and $y$ provide an accurate description 
for those filters. 

	As I did for \btvt, the zero point was calculated using inverse variance weighting to obtain
$\zpby = 0.007\pm 0.001$. For the systematic uncertainties there are two sources, which I can calculate using
the same techniques as for \btvt. From the weighted linear fit of
the data there is a 1-sigma uncertainty of 0.001 mag and from the uncertainty of 3\,000 K in the white dwarf
calibration I get 0.003 mag. Combining them in quadrature I obtain 
$\zpby = 0.007\pm 0.001$ (random) $\pm 0.003$ (systematic) magnitudes.

\subsection{\m1}

	The results for \m1\ can be seen in Fig.~\ref{byplot}, where I have chosen to represent in the $x$ axis 
the photometric \by\ instead of \m1. The reason for that choice is that \by\ is an almost monotonic function 
of the spectral slope in the optical range, which can be more easily used to detect discrepancies
between the published and the real sensitivity curves. \m1, on the other hand, is barely sensitive to
spectral slope (as it should be, since the physical variable that produces most of its variation is
metallicity). $\varepsilon_{m_1}$ was iteratively determined to be 0.007 mag. The data shows a slight
dependence with color and the slope of the corresponding weighted linear fit is $0.010\pm 0.005$ mag.
Since I have already tested the validity of the sensitivity curves for $b$ and $y$, this result
indicates that the sensitivity curve for $v$ (the other magnitude involved in \m1) may be slightly
incorrect. Note, however, that the slope is only 2 sigmas away from zero, that the variation over
the total useful \by\ range amounts only to $\pm 0.006$ magnitudes, and that no significant
differences are seen between the NGSL and Bohlin samples. Therefore, the situation is similar
to what happens for \btvt\ and I decided not to calculate any modifications in the $v$ sensitivity
curve but to include the effect as a systematic uncertainty.

\begin{figure}
\centerline{\includegraphics*[width=\linewidth]{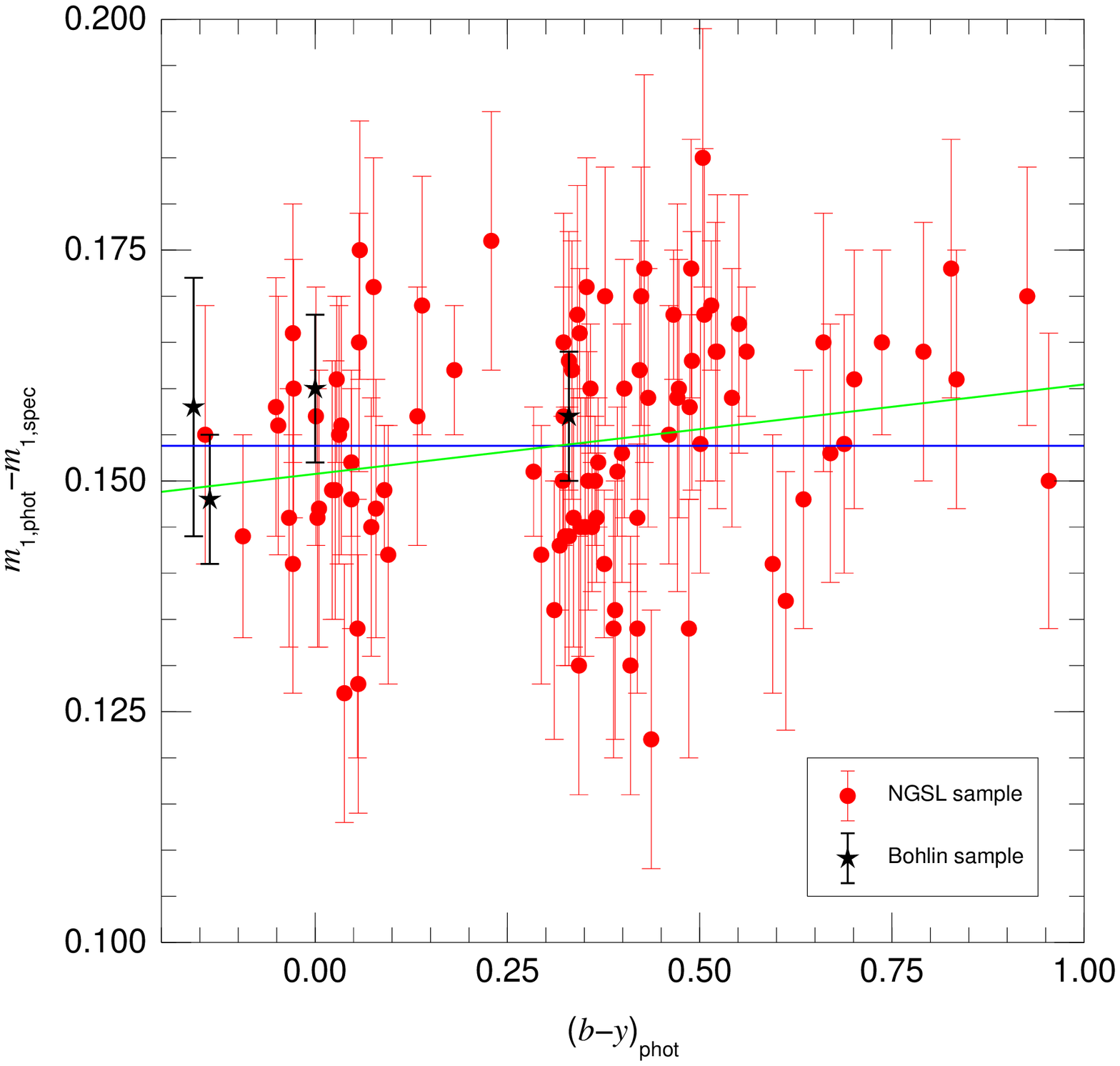}}
\caption{Comparison between photometric and spectrophotometric \m1\ indices as a function 
of photometric \by\ for the NGSL + Bohlin samples. The error bars represent the photometric
uncertainties and the horizontal blue line marks the proposed \zpm1. The green line shows the result
of a weighted linear fit to the data.}
\label{m1plot}
\end{figure}

	The zero point was calculated using inverse variance weighting to obtain 
$\zpm1 = 0.154\pm 0.001$ mag. The difference between the zero point and the linear fits yields a
1-sigma systematic uncertainty of 0.003 mag and the uncertainty in the white dwarf calibration contributes with 
less than 0.001 magnitudes, as expected from the weak dependence of \m1\ on temperature. The final result is
then $\zpm1 = 0.154\pm 0.001$ (random) $\pm 0.003$ (systematic) magnitudes.

\subsection{\c1}

\begin{figure}
\centerline{\includegraphics*[width=\linewidth]{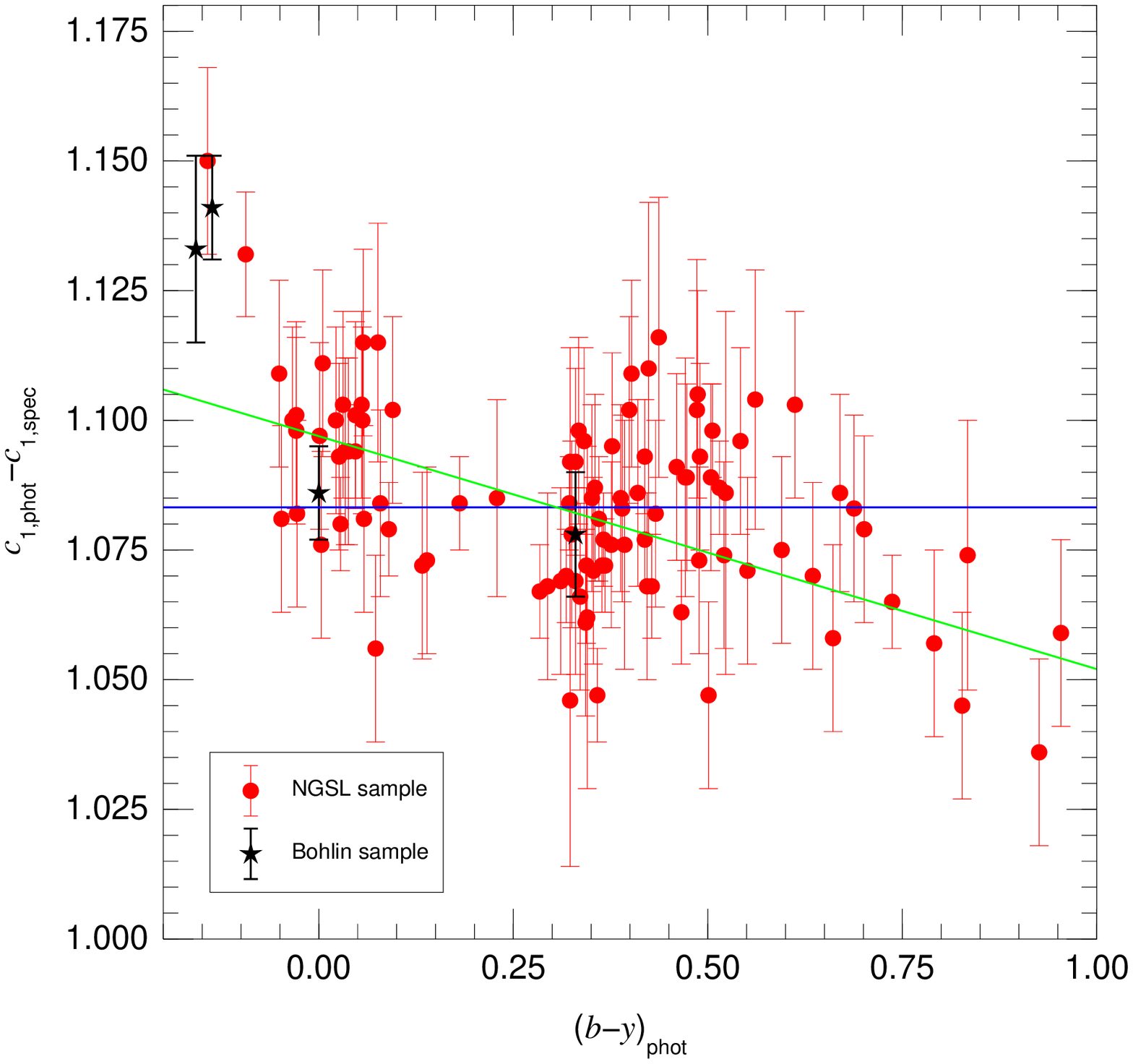}}
\caption{Comparison between photometric and spectrophotometric \c1\ indices as a function 
of photometric \by\ for the NGSL + Bohlin samples using the \citet{Mats69} sensitivity curves. The error bars 
represent the photometric uncertainties and the horizontal blue line marks the weighted mean for
the vertical coordinate. The green line shows the result of a weighted linear fit to the data.}
\label{c1oldplot}
\end{figure}

	I performed for \c1\ an analysis similar to the one for \m1\ using again the \citet{Mats69}
sensitivity curves. The $(b-y)_{\rm phot}$ vs. $c_{1,{\rm phot}}-c_{1,{\rm spec}}$ plot in
Fig.~\ref{c1oldplot} clearly shows that a color term is present, with the bluer stars above the mean
and the redder ones below it. The measured slope is $-0.045\pm 0.006$, which is 7.5 sigmas away from
zero\footnote{The uncertainty in the slope depends on the exact value of $\varepsilon_{c_1}$, which is
determined later.} and the variation over the full range is $\pm 0.027$ magnitudes. Given the results I found for
\by\ and \m1, I consider that variation too large to be tolerated as a systematic uncertainty. The likely culprit
is one of the sensitivity curves and, given that I have previously determined that the 
\citet{Mats69} $b$ filter curve is correct and that the corresponding $v$ curve 
introduces only small errors in the synthetic magnitudes, I conclude that it is the $u$ curve
definition the one that needs to be recalculated.

	A new Str\"omgren $u$ sensitivity curve was derived by $\chi^2$ minimization of 
$c_{1,{\rm phot}}-c_{1,{\rm spec}}$. I used a custom-made IDL code that 
includes the curve-fitting package written by Craig 
Markwardt\footnote{{\tt http://cow.physics.wisc.edu/~craigm/idl/idl.html}}, which allows for the
fitting of an arbitrary function using $\chi^2$-minimization with restrictions on the parameters. I
built the function in the following manner: (a) I selected 10 pivot wavelengths at 75 \AA\ intervals
between 3\,150 \AA\ and 3\,825 \AA. (b) The sensitivity at both extremes was fixed to be zero, given that we
expect the result to be only slightly different from the \citet{Mats69} curve. (c) I placed additional
constraints on the parameters to allow for a single-peaked function for the same reason. (d) This left
nine free parameters to be fit, the sensitivity at the eight intermediate pivot wavelengths and 
\zpc1. The initial guesses for the parameters were derived from the \citet{Mats69} $u$ curve. (e) The sensitivity 
curve between the pivot wavelengths was calculated using linear interpolation followed by smoothing with 
a box filter. 

\begin{figure}
\centerline{\includegraphics*[width=\linewidth]{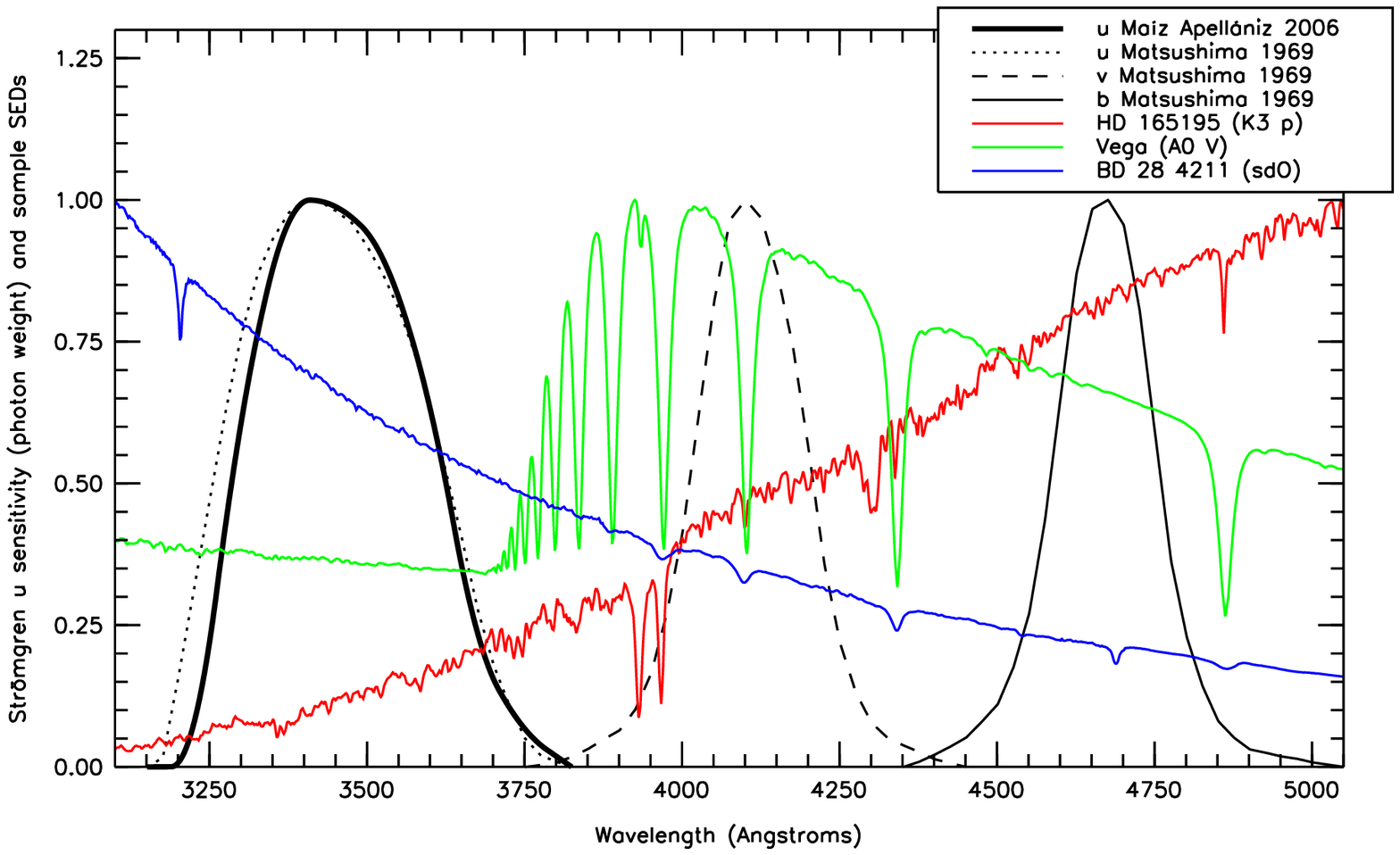}}
\caption{\citet{Mats69} photon-counting normalized sensitivity curves for the Str\"omgren $u$, $v$, and 
$b$ filters and the new curve for $u$ derived in this article. Three normalized SEDs from the two 
samples used to calibrate the Str\"omgren photometry are also shown as $f_\lambda$.}
\label{usthroughput}
\end{figure}

	An initial run of the code produced a first estimate for both the sensitivity curve and \zpm1,
but the reduced $\chi^2$ was too large due to the underestimation of $\varepsilon_{c_1}$. The code was 
run several times, changing $\varepsilon_{c_1}$ until a value of 1.0 was reached for the reduced
$\chi^2$. The sensitivity curve itself was found to change little between iterations.
$\varepsilon_{c_1}$ was determined to be 0.009 mag and \zpc1\ (which here is one of the free parameters
in the sensitivity-curve fitting procedure) to be $1.092\pm 0.002$ magnitudes. The old and new 
sensitivity curves for Str\"omgren
$u$ are shown in Fig.~\ref{usthroughput}, along with the sensitivity curves for Str\"omgren $v$ and $b$
and three selected spectra. The two curves for $u$ are found to be quite similar, with the only
exception of the short-wavelength edge, which is redder by $\approx 35$ \AA\ for the new result. 

\begin{figure}
\centerline{\includegraphics*[width=\linewidth]{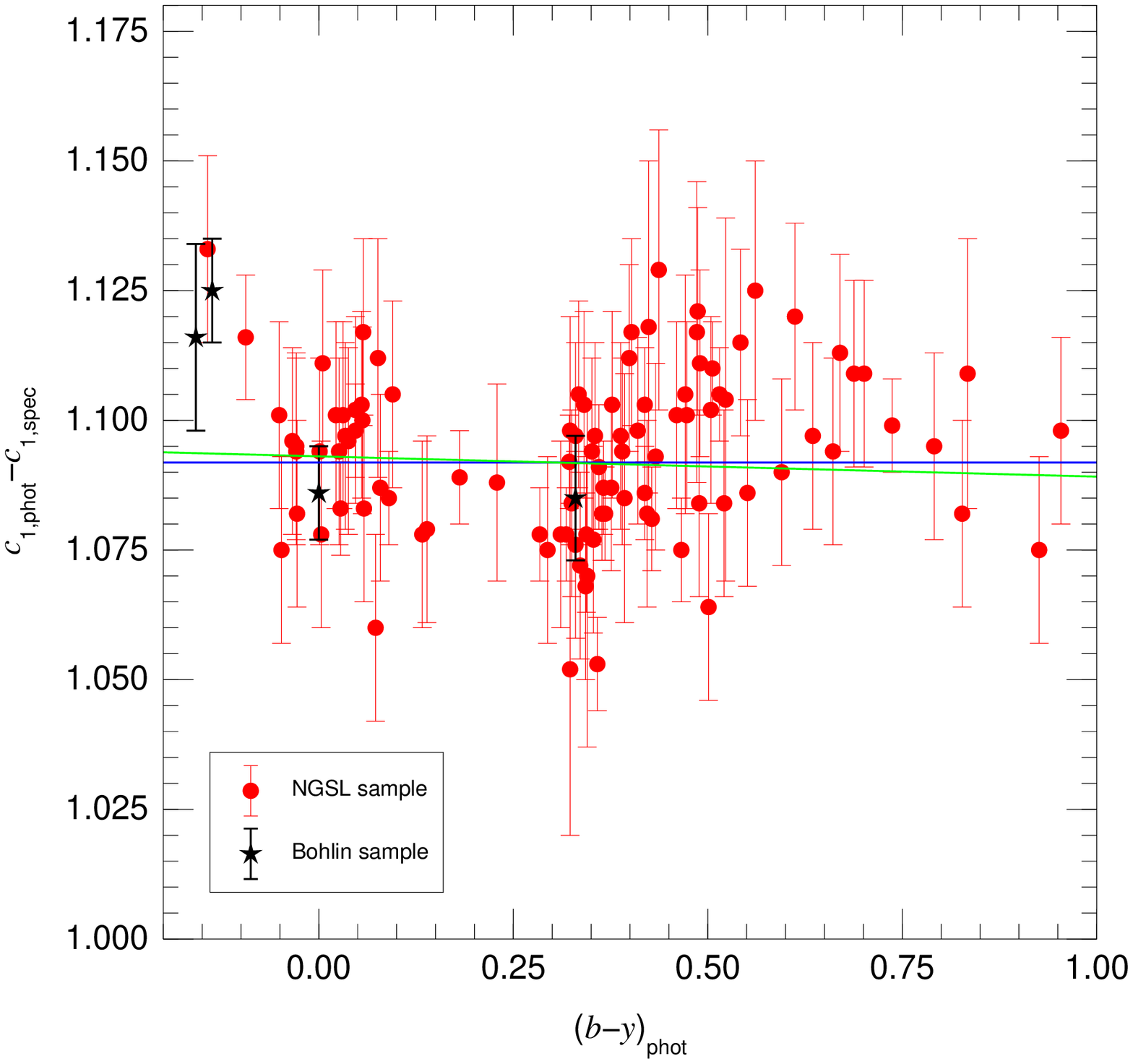}}
\caption{Comparison between photometric and spectrophotometric \c1\ indices as a function 
of photometric \by\ for the NGSL + Bohlin samples using the sensitivity curves proposed in this article. The error 
bars represent the photometric uncertainties and the horizontal blue line marks the proposed \zpc1.
The green line shows the result of a weighted linear fit to the data.}
\label{c1plot}
\end{figure}

	The new version of the $(b-y)_{\rm phot}$ vs. $c_{1,{\rm phot}}-c_{1,{\rm spec}}$ plot is 
shown in Fig.~\ref{c1plot}. As expected, the new curve does a much better job of fitting the data. The
color term is now absent, as a weighted linear fit yields a slope of $-0.004\pm 0.006$, less than 1
sigma away from zero. Again, as for \by\ and \m1, no significant differences are perceived between the
NGSL and Bohlin samples. The only outstanding issue is that the four bluest stars (two from the NGSL and
two from the Bohlin samples) are still 1-3 sigmas above the zero point. In those cases the new sensitivity 
curve is a clear improvement (previously the same stars were 3-7 sigmas above the zero point) but the 
existence of such a residual systematic effect is likely a manifestation of the inherent difficulty of doing
ground-based photometry to the left of the Balmer jump. 

	I can conclude that the new sensitivity curve does a better job of fitting the data than the 
old one (albeit not a perfect one). As for the error budget for the zero point, the slope of the fit
contributes 0.001 mag to the systematic uncertainty and the uncertainty in the temperature scale (again,
calculated from the different colors of high-gravity Kurucz models) adds 0.004 mag. Hence, I obtain
$\zpc1 = 1.092 \pm 0.002$ (random) $\pm 0.004$ (systematic) magnitudes.

\section{Johnson $UBV$}

\subsection{Description and sample selection}

	The Johnson $UBV$ standard system \citep{JohnMorg53} is the most commonly used broad-band optical photometry 
system. It covers a range of wavelengths similar to the Str\"omgren system but with only three filters. The
quantities that are presented in most references are the $V$ magnitude and the two \bv\ and \ub\ colors. Both
colors are a function mostly of temperature and extinction but metallicity and gravity effects are also present,
especially at the lower stellar temperature end. \bv\ is a rather monotonic function of temperature but for 
$T_{\rm eff} > 20\, 000$ K is almost degenerate. \ub\ is more appropriate to measure temperatures for earlier
spectral types, but the effect of the Balmer jump makes its dependence with temperature non-monotonic, especially
for high-gravity stars.

\begin{figure}
\centerline{\includegraphics*[width=\linewidth]{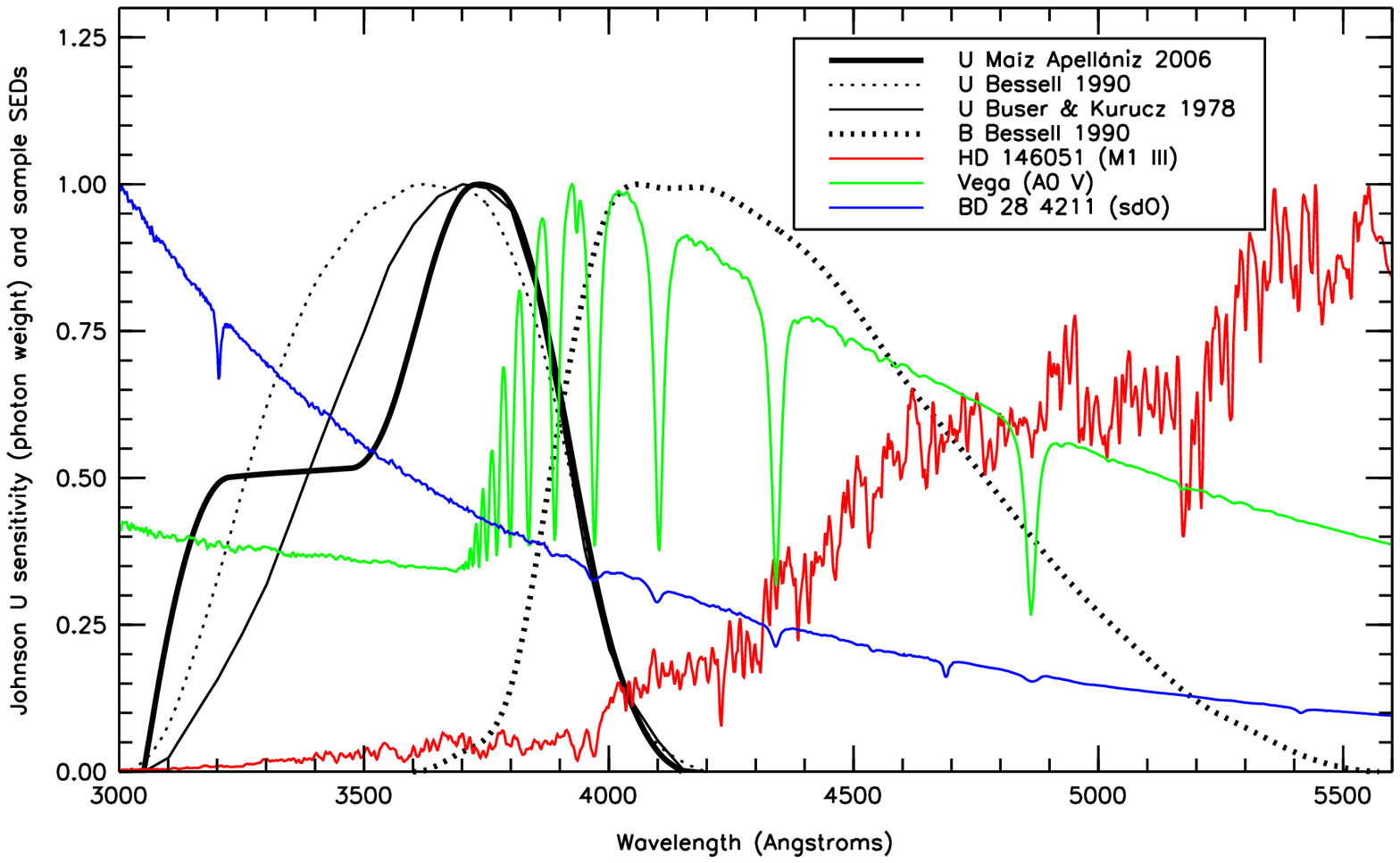}}
\caption{\citet{Bess90} $U$ (UX) and $B$ (B) and \citet{BuseKuru78} $U$ (U$_3$) photon-counting normalized 
sensitivity curves and the new curve for $U$ derived in this article. Three normalized SEDs from the two samples 
used to calibrate the Johnson photometry are also shown as $f_\lambda$.}
\label{ujthroughput}
\end{figure}

	A number of sensitivity curves for the $UBV$ system have been published over the years and a review can
be found in \citet{Bess90}. That author found a relatively good agreement between different sources for the
definitions of $B$ and $V$ but important differences in the definition of $U$. This can be seen by comparing the
sensitivity functions proposed by \citet{Bess90} with those of \citet{BuseKuru78}. The latter authors used the
definitions for $B$ and $V$ by \citet{AzusStra69} and derived a new sensitivity function for $U$. As shown in
Fig.~\ref{ujthroughput}, both sensitivity curves have similar red edges but the blue edge extends to considerably
shorter wavelengths for the \citet{Bess90} one\footnote{All sensitivity functions in Fig.~\ref{ujthroughput} and
elsewhere in this article are expressed in photon-counting form (see Appendix A).}. 

	Another important feature of both the \citet{BuseKuru78} and \citet{Bess90} studies is that two
definitions for the $B$ sensitivity function are given depending on whether they are intended for computing \ub\ or
\bv. Hence, each of those papers gives four different sensitivity functions: \citet{BuseKuru78} names them U$_3$,
B$_2$, B$_3$, and V, with $\ub = {\rm U}_3 - {\rm B}_2$ and $\bv = {\rm B}_3 - {\rm V}$, while \citet{Bess90} names
them UX, BX, B, and V, with $\ub = {\rm UX} - {\rm BX}$ and $\bv = {\rm B} - {\rm V}$. This practice can be traced
back to \citet{AzusStra69} as an attempt to correct for atmospheric extinction but it is quite obvious that it is
unphysical: $B$ is $B$, independently of whether it is used to calculate \ub\ or \bv.

	In this section I test the sensitivity curves for the Johnson $UBV$ system. In order to build the
photometric sample I used the GCPD and followed a selection procedure similar to the one in the
previous section for the Str\"omgren system. The only relevant differences are:

\begin{enumerate}
  \item I compiled one magnitude ($V$) and two colors (\bv\ and \ub). For the four white dwarfs in \citet{Bohl00}
	I used the values of $V$ in that reference and for Vega I used the value in \citet{BohlGill04a}. The rest
	of the $V$ magnitudes and all of the colors are from the GCPD.
  \item As opposed to Str\"omgren, the samples for $V$, \bv\ and \ub\ were built independently. However, the rules 
	regarding the requirement of three data points for each star and the existence of 0.035 mag cutoffs in 
	the random uncertainties were maintained.
  \item The initial values for $\varepsilon_V$, $\varepsilon_{B-V}$, and $\varepsilon_{U-B}$ were set to zero.
\end{enumerate}

	After the selection process, 108, 111, and 101 stars were present in the $V$, \bv\, and \ub\ joint
NGSL + Bohlin samples, respectively. 96 stars were present in the three samples and the 101 stars in the \ub\ 
sample were also included in the \bv\ sample. The numbers for $V$, $B$, and \bv\ including only NGSL stars are 
98, 104, and 96, respectively. The same numbers including only Bohlin stars are 10, 7, and 5, respectively.

\subsection{\bv}

\begin{figure}
\centerline{\includegraphics*[width=\linewidth]{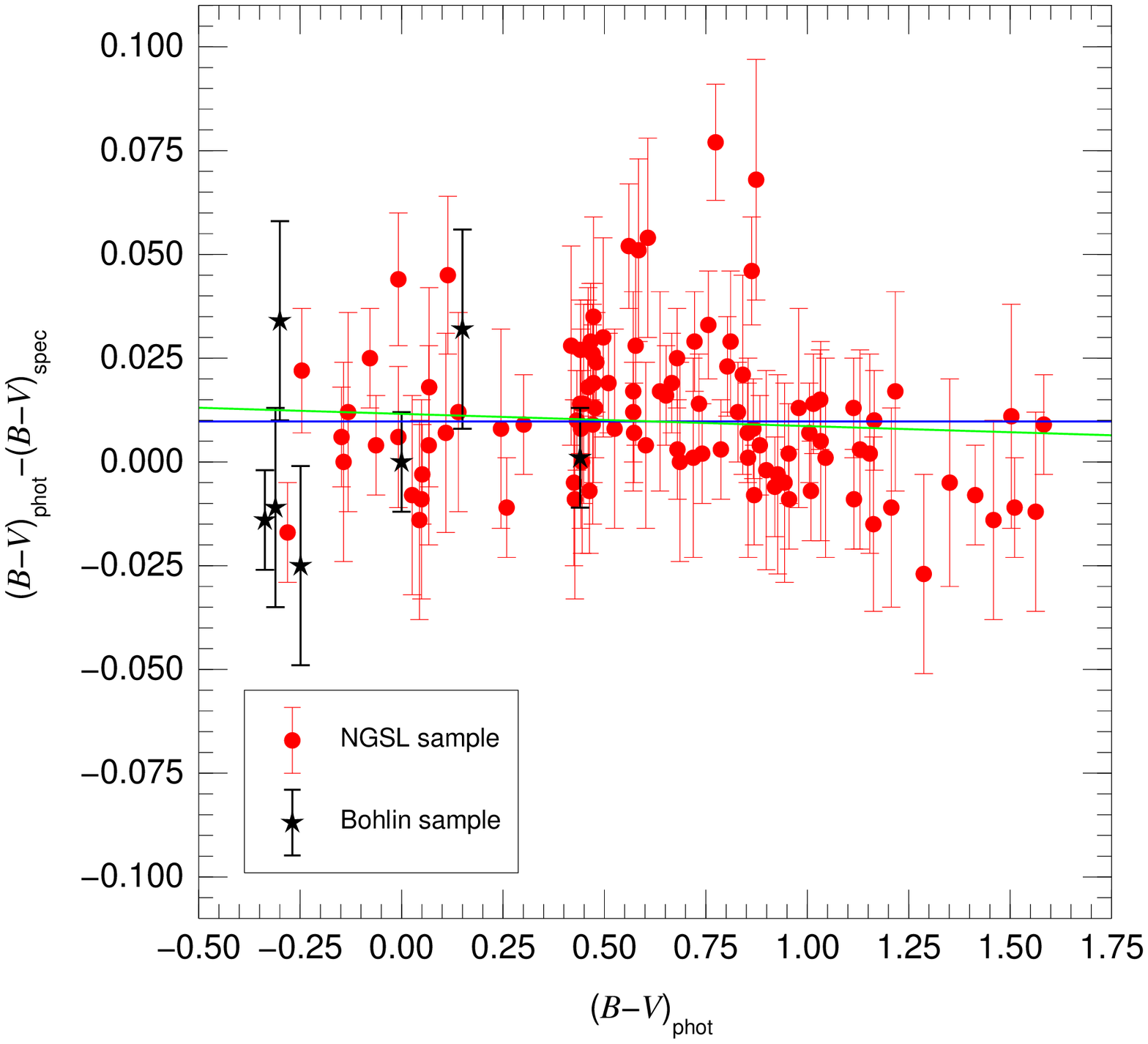}}
\caption{Comparison between photometric and spectrophotometric \bv\ colors as a function 
of photometric \bv\ for the NGSL and Bohlin \bv\ samples. The error bars represent the photometric
uncertainties and the horizontal blue line marks the proposed \zpbv. The green line shows the result
of a weighted linear fit to the data.}
\label{bvplot}
\end{figure}

	The results for \bv\ can be seen in Fig.~\ref{bvplot}. The plot uses the $B$ and $V$ sensitivity curves
of \citet{Bess90}; I tried using the \citet{BuseKuru78} and found that the results were very similar.
$\varepsilon_{B-V}$ was iteratively determined to be 0.012 mag. The data does not show any significant trend with
color, with a slope of the weighted linear fit of $-0.003\pm 0.003$, just 1 sigma away from zero. Also, no
significant difference is observed between the NGSL and Bohlin samples. Therefore, I conclude that the
\citet{Bess90} sensitivity curves for $B$ and $V$ provide an accurate description for those filters.

	The zero point was calculated using inverse variance weighting to obtain $\zpbv = 0.010\pm 0.001$ mag. For
the systematic uncertainties I obtain 0.002 mag from the slope of the weighted linear fit and 0.003 mag from the
temperature uncertainty in the white dwarf calibration. Therefore, I finally obtain 
$\zpbv = 0.010 \pm 0.001$ (random) $\pm 0.004$ (systematic) magnitudes. It is interesting to point out that the 
photometric value for Vega itself compiled from the GCPD is $\bv = 0.000 \pm 0.012$ (with the uncertainty coming 
from $\varepsilon_{B-V}$), which is less than 1 sigma away from \zpbv. This is an additional confirmation of the
consistency of the results.

\subsection{\ub}

\begin{figure}
\centerline{\includegraphics*[width=\linewidth]{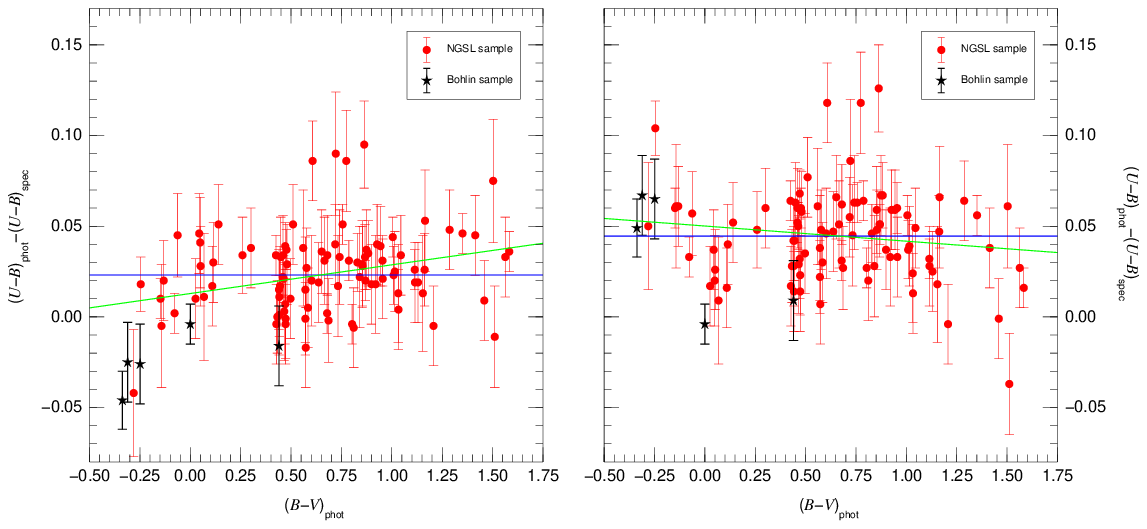}}
\caption{Comparison between photometric and spectrophotometric \ub\ colors as a function 
of photometric \bv\ for the NGSL and Bohlin \ub\ samples. The left panel uses the sensitivity curve definitions of
\citet{BuseKuru78} and the right panel those of \citet{Bess90}. The error bars represent the photometric
uncertainties and the horizontal blue line marks the weighted mean for the vertical coordinate in each panel. The 
green line shows the result of a weighted linear fit to the data in each panel.}
\label{uboldplot}
\end{figure}

	I performed for \ub\ an analysis similar to the one for \bv. Following the Str\"omgren \m1\ and \c1\ cases, 
I chose the equivalent of \by\ (i.e. \bv) as the comparison parameter since it provides a quasi-monotonic function
of the spectral slope in the optical range. Given the differences between the \citet{BuseKuru78} and \citet{Bess90}
$U$ sensitivity curves, the two cases were analyzed independently and the corresponding plots are shown in
Fig.~\ref{uboldplot}.

	The \citet{BuseKuru78} plot (left panel of Fig.~\ref{uboldplot}) shows a clear color term, with the bluer 
stars below the mean and the redder
ones above it. The measured slope is $0.016\pm 0.004$, which is indeed 4 sigmas away from zero\footnote{As it was
the case for our analysis of Str\"omgren \c1\, the exact uncertainty in the slope depends on the value of
$\varepsilon_{U-B}$, which is determined later.}, leading to a variation over the full range of $\pm 0.016$
magnitudes, which is too much to be treated simply as a systematic error. Therefore, I conclude that the
\citet{BuseKuru78} sensitivity curves do not provide an adequate representation of the literature \ub\ data.

	The \citet{Bess90} plot (right panel of Fig.~\ref{uboldplot}) also shows a color term, though weaker and of
the opposite sign as the \citet{BuseKuru78}, with the bluer stars above the mean and the redder ones below it. The
measured slope is $-0.008\pm 0.004$, which is 2 sigmas away from zero and yields a variation over the full range of
$\pm 0.008$ magnitudes. Such a variation is not very large but two additional issues exist. 
First, as previously mentioned, the 
calculation of the \ub\ color using the \citet{Bess90} definitions implies the unphysical use of a different
sensitivity curve for $B$ than the one used for the same filter in the computation of the \bv\ color. Second, the
existence of a significant slope in the weighted linear fit to the data in the right panel of Fig.~\ref{uboldplot} 
is a manifestation of a more complex effect. As we go from left to right in the plot, we see first that the OB 
stars tend to be above the mean, then the A stars tend to be below it, the F and G again above it, and finally the 
K and M stars below it. This indicates that the considered $U$ sensitivity curve is not correctly assigning weights
to the flux to the right and left of the Balmer jump. A consequence is that the derived \zpub, 0.045, is very
different from the photometric \ub\ color for Vega, the reference star (A0 V), which is $-0.004$
magnitudes. Those arguments lead me to conclude that the \citet{Bess90} sensitivity curves do not represent
the literature \ub\ data correctly, either.

	The inadequacy of the existing $U$ sensitivity curves, which I previously discussed in \cite{Maiz05c}, led 
me to derive a new one using the same procedure I followed for Str\"omgren $u$. In this case I applied $\chi^2$
minimization to the $(\ub)_{\rm phot}-(\ub)_{\rm spec}$ data and used 12 pivot wavelengths at 100 \AA\ intervals,
setting the two extremes (where the sensitivity goes to zero) at 3\,050 \AA\ and 4\,150 \AA, respectively. Also, I
eliminated the unphysical definition of an alternative $B$ filter and used the same one as for the \bv\ zero point
calculation. As it was the case for Str\"omgren $u$, several iterations with different values of 
$\varepsilon_{U-B}$ were required until a reduced $\chi^2$ of 1.0 was reached for $\varepsilon_{U-B} = 0.011$.

	The new $U$ sensitivity curve is shown in Fig.~\ref{ujthroughput}. As it happened for the Str\"omgren $u$
case, the red side of the curve is quite similar to the previous curves but the blue side is quite different. The
function shows a steep rise around 3\,100 \AA\ followed by a plateau and a new rise around 3\,600 \AA. When the
relative weights of the regions to the left and right of the Balmer jump are calculated, I find that the weight of
the region to the left is smaller than for \citet{Bess90} but larger than for \citet{BuseKuru78}. This is an
expected effect, given the opposite slope of the weighted linear fits of the two plots in Fig.~\ref{uboldplot}: a
sensitivity curve with near-zero slope should lie at an intermediate point between the two previous ones. It should 
be pointed out that during the fitting procedure I found out that the exact shape of the sensitivity curve to the 
left of 3\,600 \AA\ is poorly constrained by the data, since alternative solutions (e.g.  one with a small 
secondary peak maximum 3\,300 \AA) had similar values of $\chi^2$. This is also an expected result, since the
3\,100-3\,500 \AA\ region contains little information in the form of strong absorption lines or edges.
However, all of those alternative solutions had 
nearly constant integrated weights for the regions to the left and the right of the Balmer jump and,
therefore, produced little variation in the derived synthetic \ub. Furthermore, it is important to remember that,
as I mentioned in the introduction, it is likely that no single sensitivity curve can fit all the existing data:
the purpose of this calculation is to find the sensitivity curve that minimizes systematic errors.

\begin{figure}
\centerline{\includegraphics*[width=\linewidth]{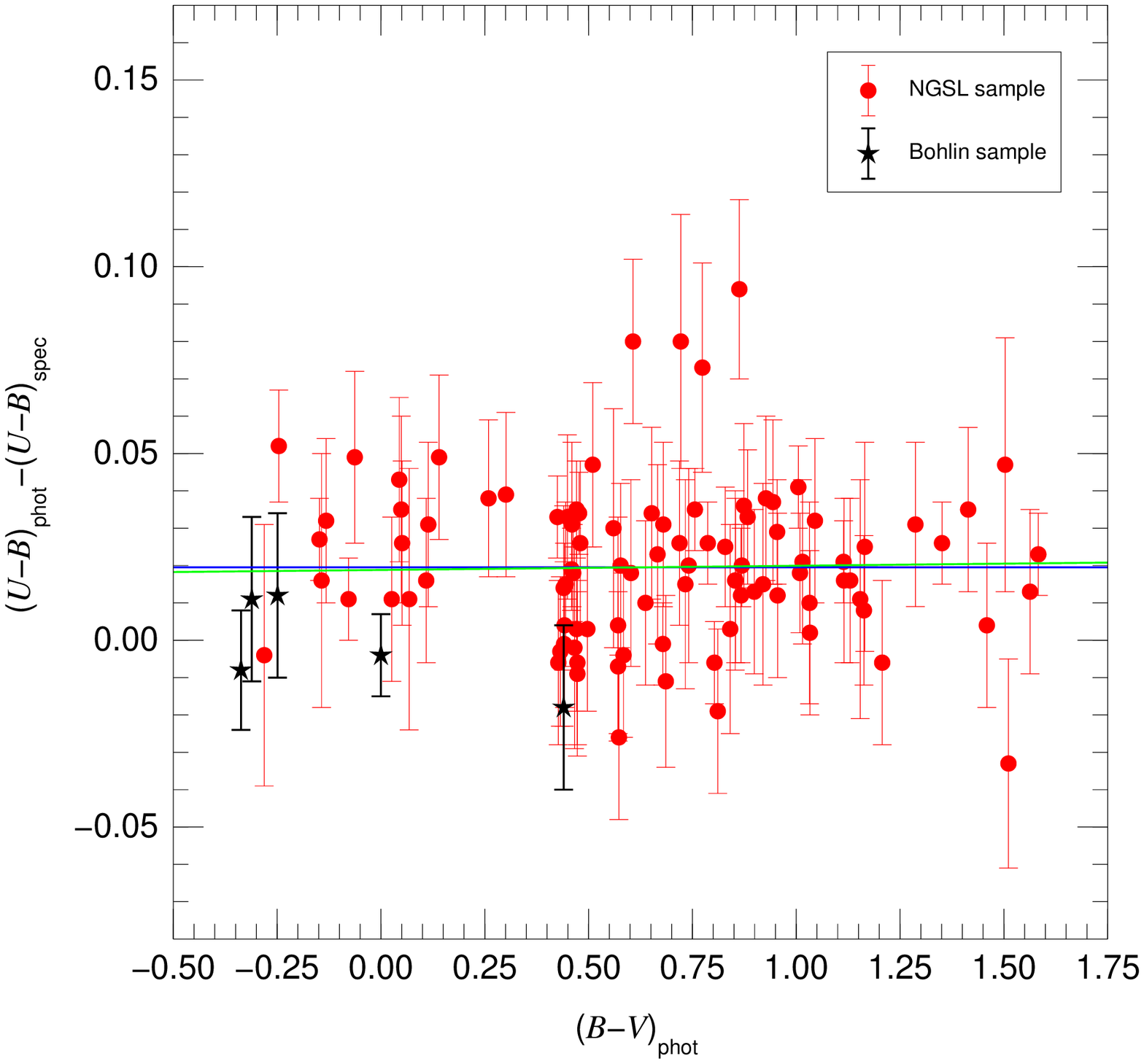}}
\caption{Comparison between photometric and spectrophotometric \ub\ colors as a function 
of photometric \bv\ for the NGSL and Bohlin \ub\ samples using the sensitivity curve definitions proposed in this
article. The error bars represent the photometric uncertainties and the horizontal blue line marks the proposed
\zpub. The green line shows the result of a weighted linear fit to the data.}
\label{ubplot}
\end{figure}

	The new version of the $(\bv)_{\rm phot}$ vs.  $(\ub)_{\rm phot}-(\ub)_{\rm spec}$ is shown in
Fig.~\ref{ubplot}. As expected, the color term is absent, with the slope of the weighted linear fit being 
$0.001\pm 0.004$. The value for \zpub, obtained from the sensitivity-curve $\chi^2$ minimization procedure, is
$0.020\pm 0.006$ magnitudes. The systematic uncertainty from 
the fitted slope is very small ($< 0.001$ mag) and the one
from the uncertainty in the white dwarf temperature scale is 0.007 magnitudes. \zpub\ is now considerably closer to
the measured photometric \ub\ color of Vega but, for $\varepsilon_{U-B} = 0.011$, is still 2 sigmas away. Also, it
should be noted that the five stars in the Bohlin sample appear to be shifted downward by $\approx 0.02$ magnitudes
with respect to the mean locus of the NGSL sample in Fig.~\ref{ubplot}. It is unlikely that the source of the 
relative displacement is due to a problem with an incorrect relative calibration of spectrophotometry of the NGSL 
sample with respect to the Bohlin sample (as it was the case for Tycho-2 in \citealt{Maiz05b}) since such a 
displacement should also show up in Fig.~\ref{c1plot}, given the similar wavelengths sampled by $U$ and $u$. The
effect could simply be a case of small number statistics (the probability that all of five points in a Gaussian
distribution lie at the same side of the mean is 1/16) but, to be conservative, I suggest that the systematic 
uncertainty be doubled to take into account the possibility that the difference between the results for the NGSL
and the Bohlin samples is real. Therefore, the proposed value for 
\zpub\ is $0.020\pm 0.006$ (random) $\pm 0.014$ (systematic) magnitudes.

\subsection{$V$}

	Up until now I have dealt with the calibration of colors and indices but not of magnitudes themselves. In
the first case the accuracy of the calibration depends ultimately on the parameters (fundamentally temperature) of 
the spectrophotometric standards and on the stability and knowledge of the observational photometric conditions.
The same is true for magnitudes but there one also runs into the spectrophotometric absolute flux calibration. 
That issue has been tackled by \citet{Bohl00} and \citet{BohlGill04a}, who combined the 5\,556 \AA\ 
absolute flux of Vega of \citet{Mege95}, Landolt $V$ photometry for the three white dwarfs used as primary 
calibrators, and STIS spectrophotometry to derive a $V$ value for Vega of $0.026\pm 0.008$ magnitudes. 

\begin{figure}
\centerline{\includegraphics*[width=\linewidth]{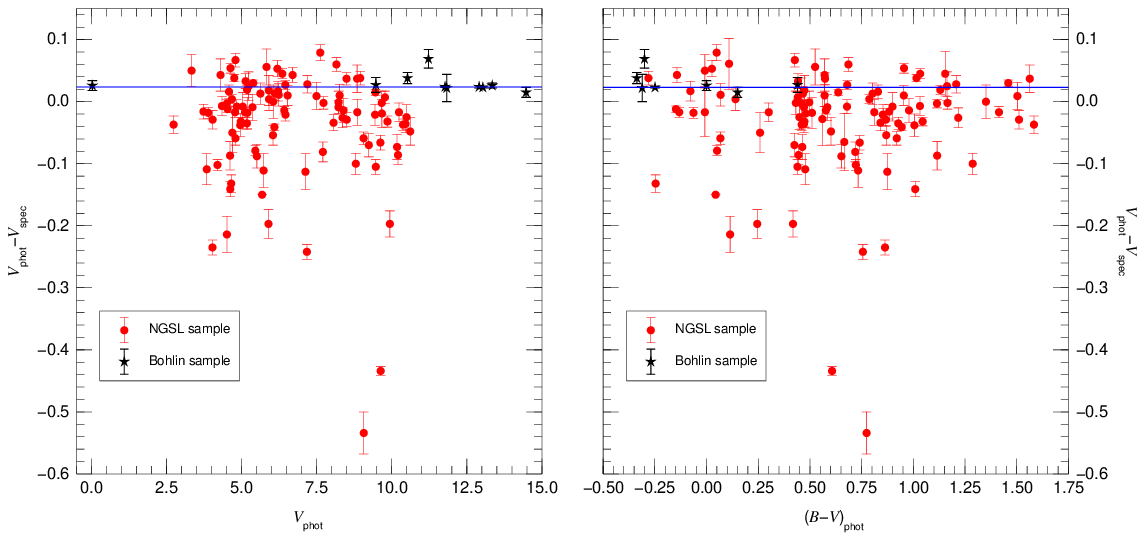}}
\caption{Comparison between photometric and spectrophotometric $V$ magnitudes as a function of 
the photometric values for $V$ (left) and \bv\ (right) for the NGSL + Bohlin samples. The error bars 
represent the photometric uncertainties and the horizontal line marks the proposed 
\zpv. The asymmetric scatter of values below the line is
due to light loss at the slit due to poor centering.}
\label{vplot}
\end{figure}

	Trying to use the Bohlin sample to derive a value for \zpv\ using the same procedure as for e.g.  \bv\ would be 
falling into a logical inconsistency: the $V$ photometry of the white dwarf calibrators and their HST spectrophotometry 
are not independent, since the absolute calibration of the second has been derived using the first as input. 
Therefore, the fact that the $V_{\rm phot}-V_{\rm spec}$ values for the Bohlin sample in Fig.~\ref{vplot} show no trend 
with $V_{\rm phot}$ or $(\bv)_{\rm phot}$ is simply a verification that the spectrophotometric calibration is
self-consistent. A similar argument takes place with the value of \zpv. In the Bohlin sample there are five stars 
where the $V$ magnitudes are taken from the two references above plus another four stars with accurate photometry 
in the GCPD (however, only six stars have also accurate \bv\ photometric colors in our data). The weighted mean of 
$V_{\rm phot}-V_{\rm spec}$ yields 0.024, which is very close to the $V$ magnitude of Vega\footnote{The reason why
it is not identical is the inclusion of the five additional stars with GCPD photometry.}. And what about using the 
NGSL sample to calculate \zpv? Unfortunately, there one runs into the same problem as for 
\zpvt\ or \zpbt\ in Paper I: the NGSL data
cannot be used to directly calculate the zero point of a magnitude because they were obtained with the 52X0.2 slit,
so significant light losses due to miscenterings are to be expected. Indeed, that effect is clearly seen in 
Fig.~\ref{vplot} in a similar manner to what happens in Fig.~4 in Paper I. Therefore, the most appropriate choice 
for \zpv\ is the $V$ magnitude for Vega, $0.026\pm 0.008$, derived by \citet{BohlGill04a}. 

\subsection{A new look at the absolute calibration of Tycho-2 photometry}

	In Paper I the \vt\ filter was calibrated through the \bt\ photometry for the Bohlin sample and the 
\zpbtvt\ derived from the combination of the NGSL and Bohlin samples. Since the last term is now known to have been
in error, a new value for \zpvt\ needs to be calculated. If I use the four stars in the Bohlin sample with
accurate \vt\ photometry and calculate the mean of $\vt_{\rm phot} - \vt_{\rm spec}$ using inverse variance weights 
I obtain $0.039\pm 0.008$ magnitudes (see right panel of Fig.~4 in Paper I). Note that the uncertainty originates
in the Tycho-2 photometry, not in the the Vega $V$ magnitude. It is interesting to notice that if the absolute
HST calibration were to be off by a multiplicative constant (let's say 1\%), it would not affect that
measurement of \zpvt\ because the zero point is measured by comparing two spectrophotometric fluxes on the same
scale (e.g. the multiplicative constant would affect both the numerator and the denominator in 
Eq.~\ref{photon})\footnote{This is the same reason why the 0.7\% uncertainty in the absolute flux of Vega 
\citep{Mege95} is not included in the error balance for \zpv.}. A different issue, of course, would be the absolute 
zero point as defined by Eq.~\ref{absrel}. 

\begin{figure}
\centerline{\includegraphics*[width=\linewidth]{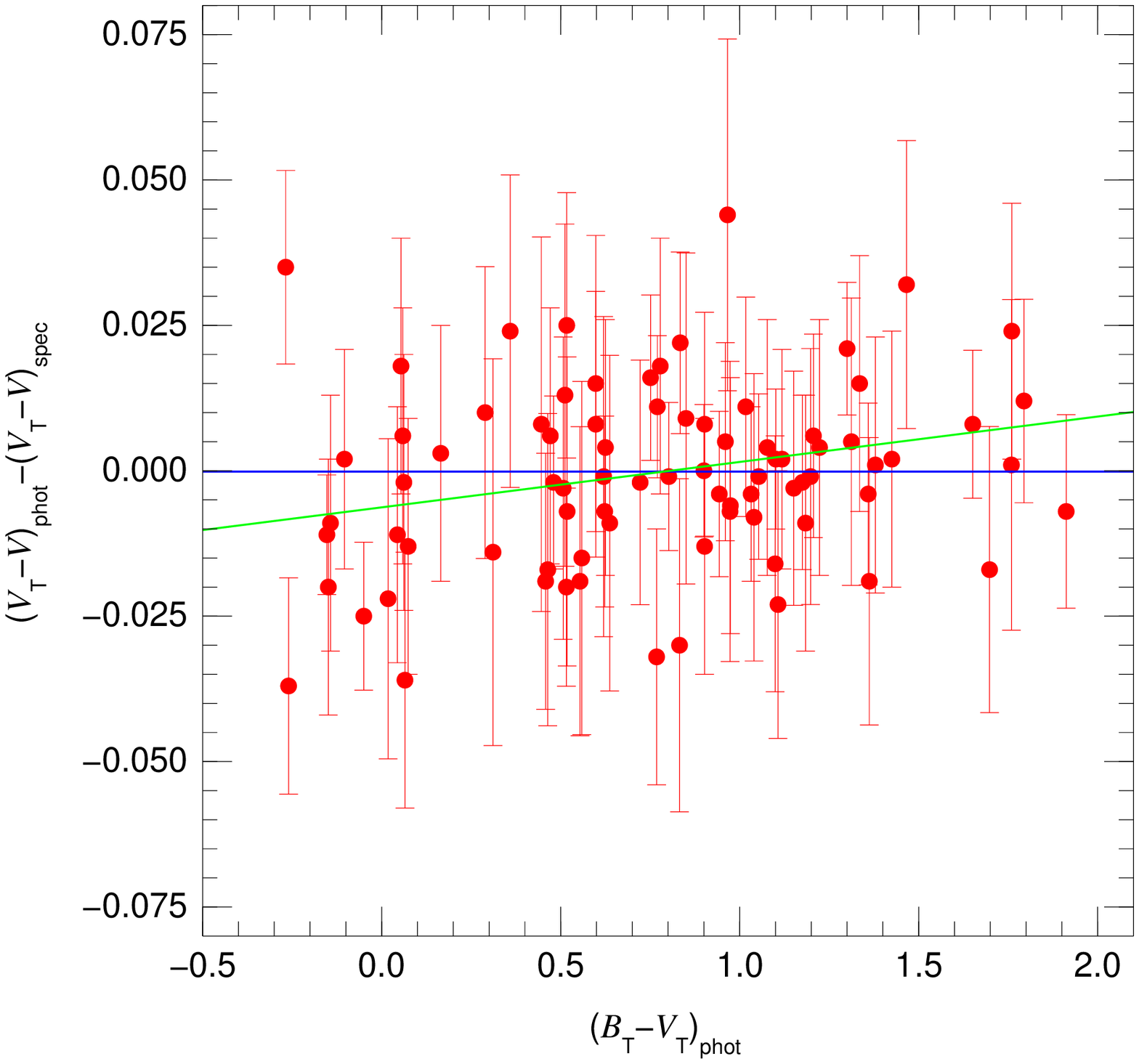}}
\caption{Comparison between photometric and spectrophotometric \tj\ magnitudes as a function of 
the photometric values \btvt\ (right) for the NGSL sample. The error bars 
represent the photometric uncertainties and the horizontal line marks the proposed 
\zptj. The green line shows the result of a weighted linear fit to the data.}
\label{tjplot}
\end{figure}

	The availability of the Johnson $V$ photometry allows me to perform an independent measurement of \zpvt.
\vt\ and $V$ measure the flux in a similar wavelength range \citep{Bess00} 
but are also different enough that \tj\ varies by
$\approx 0.26$ magnitudes as we move along the unreddened main sequence. In any case, since the \vt\ and $V$
photometric measurements are independent, I can compile photometric \tj\ values and compare them with the computed
spectrophotometric ones to derive \zptj. The results are shown in Fig.~\ref{tjplot}. There is a color term present 
but the slope is quite small, $0.008\pm 0.004$, and is only 2 sigmas away from zero so, as I have done for other
colors, the effect can be simply included as a systematic uncertainty. Using inverse variance weights I obtain 
$\zptj = 0.000\pm 0.002$ magnitudes. For the systematic uncertainty there are 0.005 magnitudes due to the color term 
and 0.001 magnitudes due to the uncertainty in the white dwarf temperature scale, leading to a final value of
$\zptj = 0.000 \pm 0.002$ (random) $\pm 0.005$ (systematic) magnitudes. \zpvt\ can now be obtained by adding the
\zpv\ value from the previous subsection.  Merging random and systematic uncertainties for simplicity, I get 
$\zpvt = 0.026 \pm 0.010$. 

	The measurements of \zpvt\ in the two previous paragraph are independent and compatible, as they are just 
one sigma away ($[0.039-0.026]/\sqrt{0.008^2+0.010^2} = 1.0$). Furthermore, since there are no systematic 
uncertainties in common in the two derivations, they can be combined using inverse variance weights to obtain a 
best value for \zpvt\ of $0.034\pm 0.006$.

	The value for \zpbt\ derived in Paper I from the Bohlin sample is $0.078\pm 0.009$. If I combine this with
the new value for \zpvt\ (derived from the NGSL \tj\ colors and from \citealt{BohlGill04a}) I obtain an
alternative measurement for \zpbtvt\ of $0.044\pm 0.011$ magnitudes, which is consistent with the result of 
$0.033\pm 0.001$ (random) $\pm 0.005$ (systematic) previously derived, thus providing an additional check on the
validity of the zero points.

\section{Summary and applications}

	I have analyzed the sensitivity curves for the Tycho-2, Str\"omgren, and Johnson photometric systems and have
determined that for 7 of the 9 curves there are values in the literature consistent with the available photometry and HST 
spectrophotometry. The recommended photon-counting sensitivity curves are given in 
Tables~\ref{tthroughputs},~\ref{sthroughputs},~and~\ref{jthroughputs}. The results for \bt\ and \vt\ are taken from
\citet{Bess00}, for Str\"omgren $vby$ from \citet{Mats69}, and for Johnson $BV$ from \citet{Bess90} and have been converted
in all cases into photon counting curves. The results for Str\"omgren $u$ and Johnson $U$ have been derived in this paper
after finding that the published curves do not provide consistent results with both the photometry and the
spectrophotometry. For those two cases, the differences between the literature and the new curves reside on 
the blue edge of a filter centered around 3\,400-3\,700 \AA, which indicates that the likely culprit is an 
incorrect characterization of atmospheric extinction.
	
	I show in Tables~\ref{colorzp}~and~\ref{magnitudezp} the recommended zero points for the colors, indices, and
magnitudes calculated in this article. All zero points use Vega as the reference spectrum. In Table~\ref{magnitudezp} I
also include the zero points for $y$ computed by \citet{HolbBerg06}\footnote{I could not compute \zpby\ from the data in
this paper because the GCPD provides Johnson $V$ magnitudes instead of Str\"omgren $y$ magnitudes and a small color term 
$y-V$ exists, just as for \tj.} and for the 2MASS system computed by \citet{Coheetal03}.

	Recently, \citet{HolbBerg06} have combined optical/IR photometry and SED models of very low extinction DA white 
dwarfs with the \citet{Bohl00} HST flux calibration to derive results similar to the ones in this paper. Their color/index 
zero points for the Str\"omgren and Johnson systems are independent from the ones here because they are based on a largely 
different sample and because they use model SEDs instead of observed ones. Their values for \zpby, \zpm1, and \zpbv\ are
0.004, 0.154, and 0.007 magnitudes, respectively, all of which agree within 1 sigma with the ones in this paper. Since those
authors do not attempt to fit Str\"omgren $u$ or Johnson $U$ sensitivity curves, it is not possible to compare their zero 
points for \c1\ and \ub\ directly with mine. However, the fact that their Str\" omgren data show the highest scatter for the
$u$ band and that their Johnson filter with the largest dispersion is $U$ points in the same direction as my conclusion that
new sensitivity curves are required in those two cases. Therefore, the results in the two articles are consistent.

\begin{figure}
\centerline{\includegraphics*[width=0.31\linewidth]{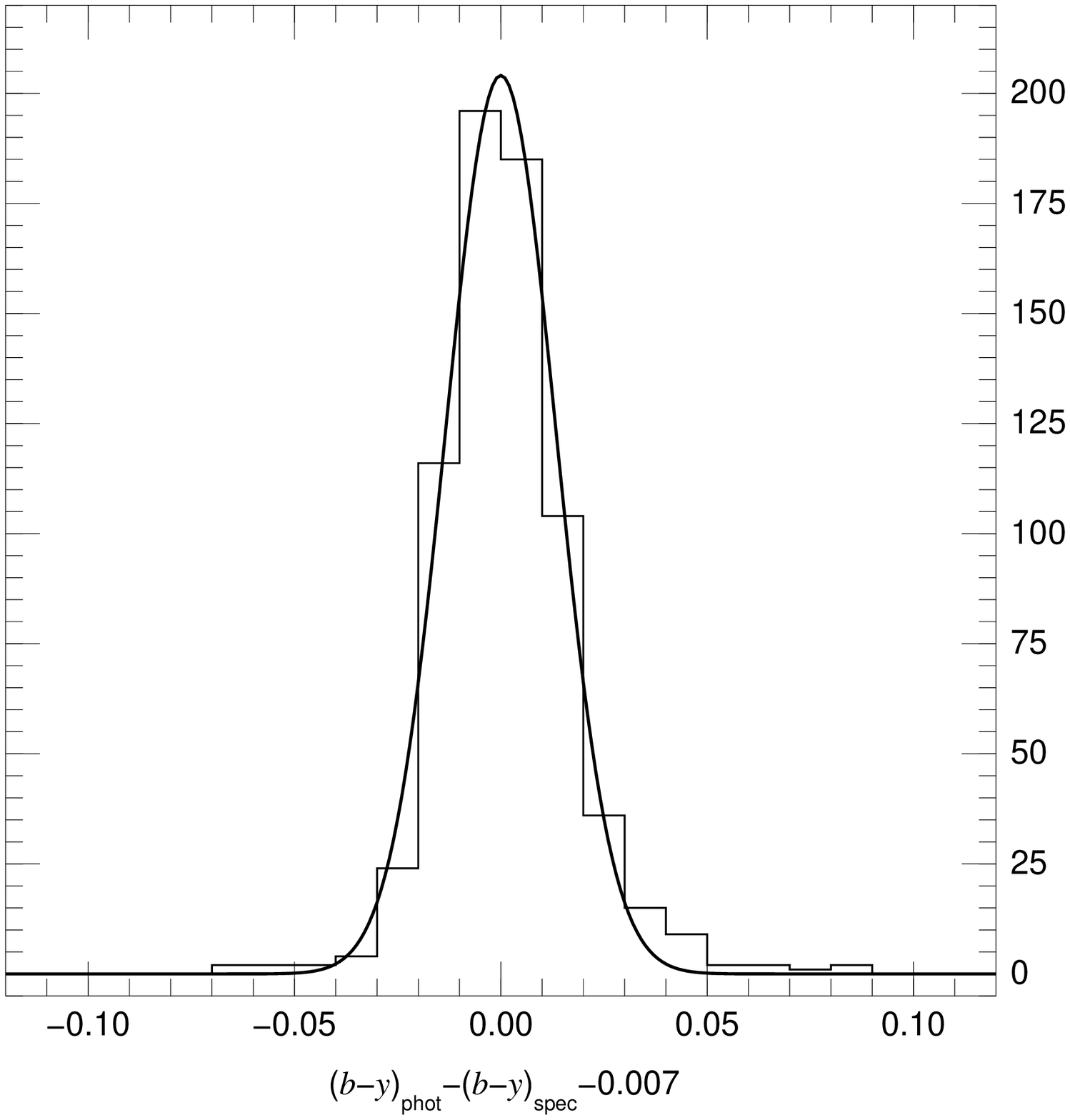} \
            \includegraphics*[width=0.31\linewidth]{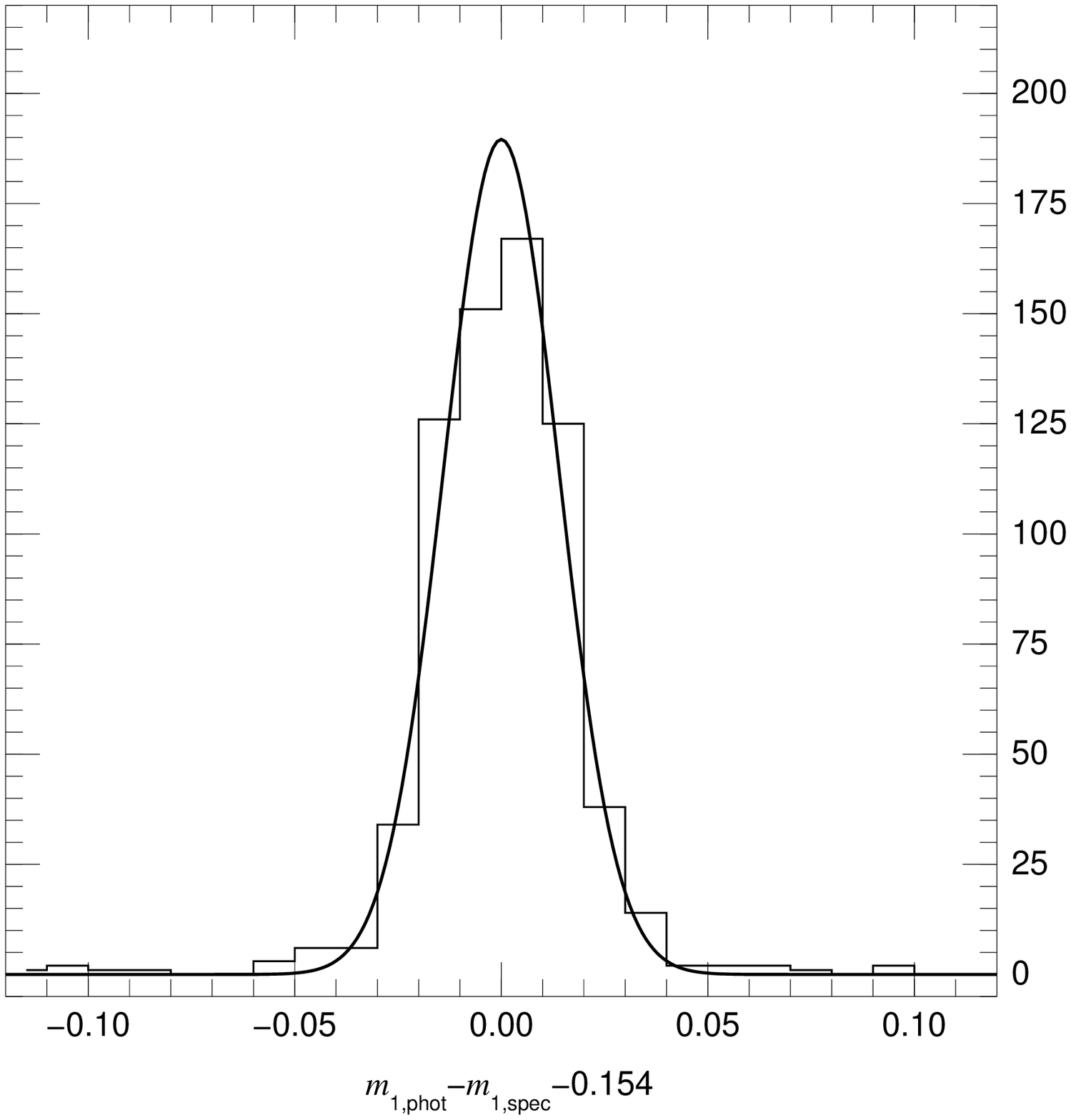} \
            \includegraphics*[width=0.31\linewidth]{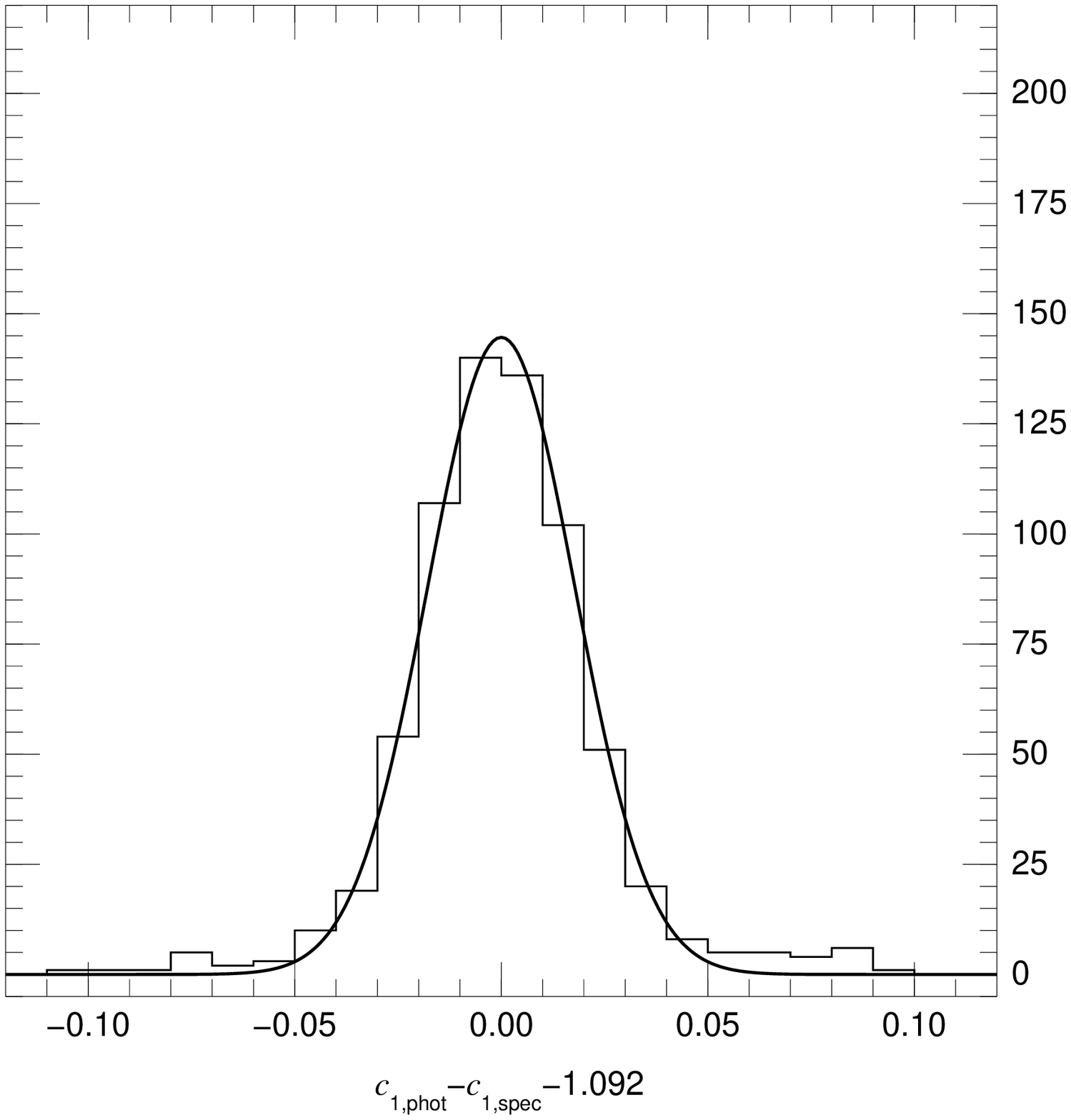}}
\caption{Comparison between the photometric and spectrophotometric Str\"omgren values for the individual data 
points in the GCPD. The left, center, and right panels show \by\, \m1, and \c1, respectively. Gaussians with zero mean and
$\sigma$ of 0.013, 0.014, and 0.018 magnitudes, respectively, are plotted for comparison.}
\label{stromgrenhisto}
\end{figure}

\begin{figure}
\centerline{\includegraphics*[width=0.47\linewidth]{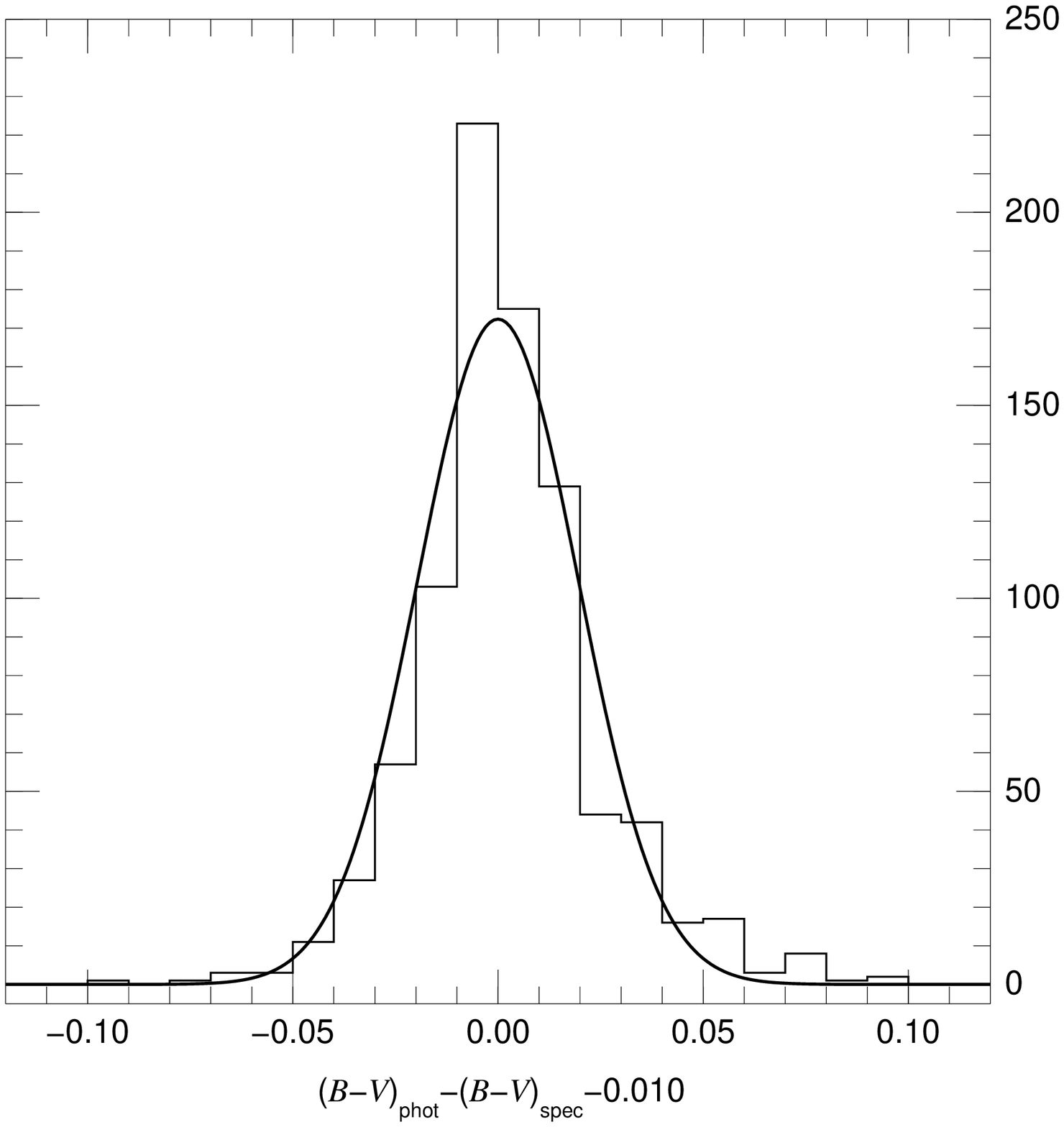} \
            \includegraphics*[width=0.47\linewidth]{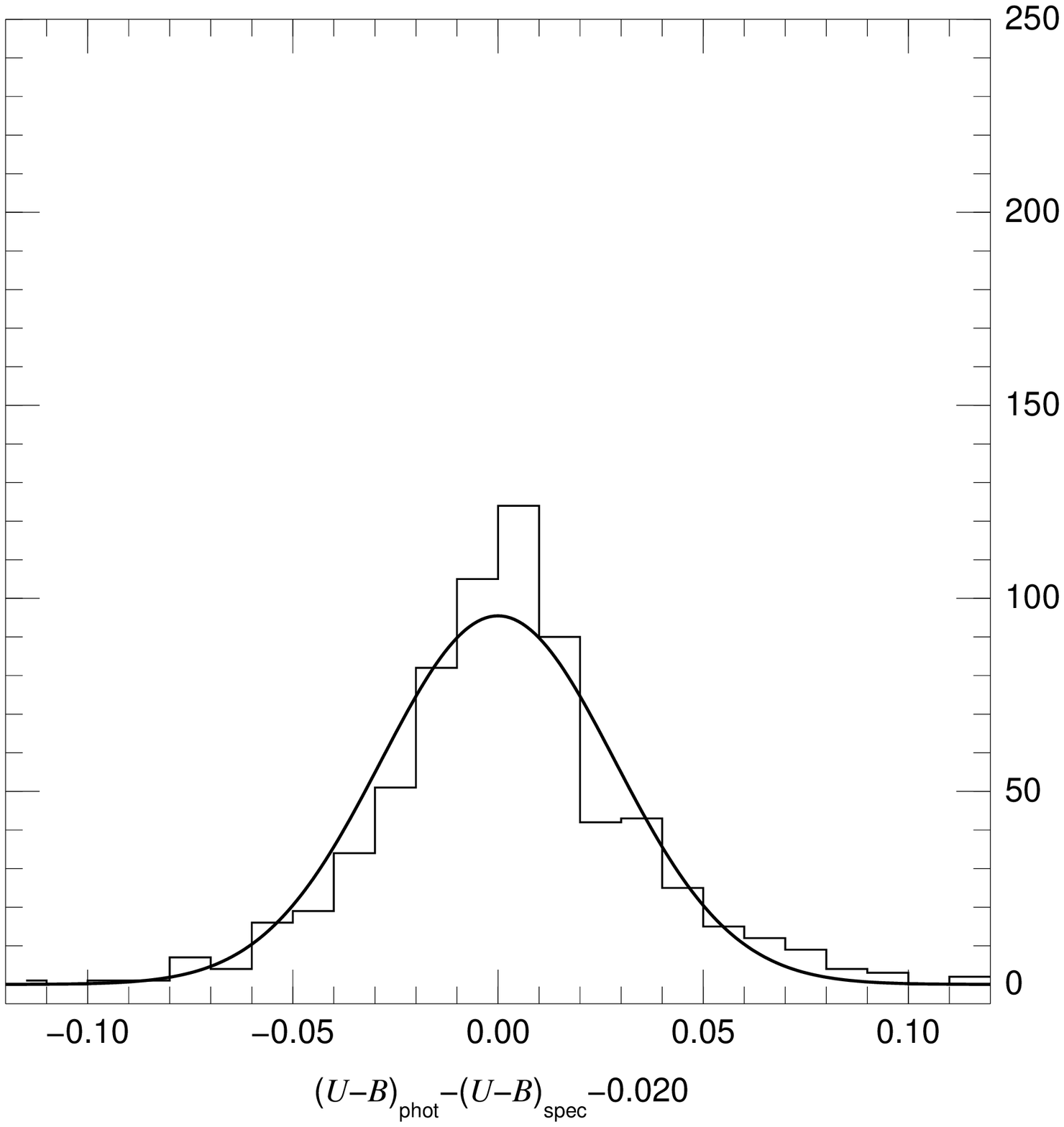}}
\caption{Comparison between the photometric and spectrophotometric Johnson values for the individual data 
points in the GCPD. The left and right panels show \bv\ and \ub, respectively. Gaussians with zero mean and
$\sigma$ of 0.020 and 0.028 magnitudes, respectively, are plotted for comparison.}
\label{johnsonhisto}
\end{figure}

	The most direct application of this work is the use of the zero points in Tables~\ref{colorzp}~and~\ref{magnitudezp}
and of the two new sensitivity curves for synthetic photometry. The latest version of CHORIZOS \citep{Maiz04c} at the time
of this publication (2.0) incorporates them. Another application is to provide 
uncertainties for the Str\"omgren and Johnson colors and indices in the GCPD. To do the latter, I start with the
photometry originally compiled in sections~4~and~5. For each data point for each star in the NGSL sample I can now compute 
the difference between the photometric and spectrophotometric values for each color and index and obtain the corresponding 
distributions. The total number of data points for each of the five colors/indices turns out to be between 662 and 871, much 
larger than the numbers used to test the sensitivity curves because now all stars in the sample are used and no average 
is done for each star; in this way one can obtain an estimate for the typical uncertainty associated with a literature
value. The distributions are shown in Figs.~\ref{stromgrenhisto}~and~\ref{johnsonhisto} and can be well approximated by a
Gaussian plus tails. The tails are defined as being more than 3.5 sigmas away from the mean, with the standard
deviation calculated iteratively. The value of 3.5 sigmas is chosen because $1/[1-{\rm errorf}(3.5/\sqrt{2})] = 2\,149$,
implying that a sample of roughly three times the ones we have here are required for at least one data point to be 
at larger distances from the mean. For each color/index I find that 2-6\% of the data points are in the tails and I
take those objects to be either misidentifications or cases with incorrect data reductions. Hence, they are not considered
for the subsequent analysis.

	The mean for the distributions of $(\by)_{\rm phot}-(\by)_{\rm spec}$, $\m1_{\rm phot}-\m1_{\rm spec}$, and
$\c1_{\rm phot}-\c1_{\rm spec}$ are all within 0.001 magnitudes of the corresponding zero
points. The dispersions, which can be taken to be the estimates for the uncertainty associated with a single data point in
the GCPD, are 0.013, 0.014, and 0.018 magnitudes, respectively. These results provide an additional check to the validity of
the sensitivity curves and zero points calculated in this article\footnote{Also note that the dispersions are approximately 
equal to $2\varepsilon_X$ in each case, thus confirming that our previous uncertainty estimates were reasonable.}
and show that the typical precision for a single value of a Str\"omgren color or
index in the literature is $1-2$\%. In Tables~\ref{bysigmas},~\ref{m1sigmas},~and~\ref{c1sigmas} the dispersions are given
as a function of $V$ magnitude and $N$, the number of measurements used for an individual data point \citep{HaucMerm98}. The
variation with $V$ and $N$ is in the expected direction (brighter stars or those with more measurements have lower
dispersions) but the effect is very small, so using a single value for the photometric uncertainty should not be a bad
approximation.

	The mean for the distributions of $(\bv)_{\rm phot}-(\bv)_{\rm spec}$ and $(\ub)_{\rm phot}-(\ub)_{\rm spec}$
are 0.0016 and 0.0022 magnitudes larger than the respective zero points, slightly larger than for the Str\"omgren cases but
still within the values expected from the uncertainties derived for \zpbv\ and \zpub. The associated dispersions are 0.020
and 0.028 magnitudes\footnote{Once again, not too different from $2\varepsilon_X$.}, 
also slightly larger than in the Str\"omgren cases. The results in
Tables~\ref{bvsigmas}~and~\ref{ubsigmas} show a weak dependency of the dispersions with $V$ and a negligible dependence with
$N$, so a single value for the photometric uncertainty should also be appropriate for Johnson photometry. I conclude that
the typical precision for \bv\ and \ub\ values in the literature is in the $2-3$\% range, slightly larger than for
Str\"omgren.

\acknowledgments

I would like to thank 
Ralph Bohlin and an anonymous referee for useful comments on this paper, 
Jay Holberg for granting me access to his results prior to publication and for useful suggestions, 
Steve Willner for bringing into my attention an issue with the 2MASS zero points,
and Charles Proffitt and the rest of the STIS team at STScI for their help with the spectrophotometric calibration.

\bibliographystyle{apj}
\bibliography{general}

\begin{thebibliography}{32}
\expandafter\ifx\csname natexlab\endcsname\relax\def\natexlab#1{#1}\fi

\bibitem[{{A{\v z}usienis} \& {Strai{\v z}ys}(1969)}]{AzusStra69}
{A{\v z}usienis}, A., \& {Strai{\v z}ys}, V. 1969, Soviet Astronomy, 13, 316

\bibitem[{{Bessell}(1990)}]{Bess90}
{Bessell}, M.~S. 1990, PASP, 102, 1181

\bibitem[{{Bessell}(2000)}]{Bess00}
---. 2000, PASP, 112, 961

\bibitem[{{Bessell} {et~al.}(1998){Bessell}, {Castelli}, \&
  {Plez}}]{Bessetal98}
{Bessell}, M.~S., {Castelli}, F., \& {Plez}, B. 1998, A\&A, 333, 231

\bibitem[{Bianchi {et~al.}(1999)}]{Bianetal99}
Bianchi, L., {et~al.} 1999, Mem. Soc. Astron. Ital., 70, 365

\bibitem[{{Bohlin}(2000)}]{Bohl00}
{Bohlin}, R.~C. 2000, AJ, 120, 437

\bibitem[{{Bohlin} {et~al.}(2001){Bohlin}, {Dickinson}, \&
  {Calzetti}}]{Bohletal01}
{Bohlin}, R.~C., {Dickinson}, M.~E., \& {Calzetti}, D. 2001, AJ, 122, 2118

\bibitem[{{Bohlin} \& {Gilliland}(2004{\natexlab{a}})}]{BohlGill04b}
{Bohlin}, R.~C., \& {Gilliland}, R.~L. 2004{\natexlab{a}}, AJ, 128, 3053

\bibitem[{{Bohlin} \& {Gilliland}(2004{\natexlab{b}})}]{BohlGill04a}
---. 2004{\natexlab{b}}, AJ, 127, 3508

\bibitem[{{Buser} \& {Kurucz}(1978)}]{BuseKuru78}
{Buser}, R., \& {Kurucz}, R.~L. 1978, A\&A, 70, 555

\bibitem[{Cohen {et~al.}(2003)Cohen, Wheaton, \& Megeath}]{Coheetal03}
Cohen, M., Wheaton, W.~A., \& Megeath, S.~T. 2003, AJ, 126, 1090

\bibitem[{ESA(1997)}]{ESA97}
ESA. 1997, {The Hipparcos and Tycho Catalogues} (ESA SP-1200)

\bibitem[{{Gregg} {et~al.}(2004){Gregg}, {Silva}, {Rayner}, {Valdes},
  {Worthey}, {Pickles}, {Rose}, {Vacca}, \& {Carney}}]{Gregetal04}
{Gregg}, M.~D., {Silva}, D., {Rayner}, J., {Valdes}, F., {Worthey}, G.,
  {Pickles}, A., {Rose}, J.~A., {Vacca}, W., \& {Carney}, B. 2004, American
  Astronomical Society Meeting Abstracts, 205

\bibitem[{{Hauck} \& {Mermilliod}(1998)}]{HaucMerm98}
{Hauck}, B., \& {Mermilliod}, M. 1998, A\&AS, 129, 431

\bibitem[{Holberg \& Bergeron(2006)}]{HolbBerg06}
Holberg, J.~B., \& Bergeron, P. 2006, AJ (submitted)

\bibitem[{{Johnson} \& {Morgan}(1953)}]{JohnMorg53}
{Johnson}, H.~L., \& {Morgan}, W.~W. 1953, ApJ, 117, 313

\bibitem[{{Kimble} {et~al.}(2000){Kimble}, {Goudfrooij}, \&
  {Gilliland}}]{Kimbetal00}
{Kimble}, R.~A., {Goudfrooij}, P., \& {Gilliland}, R.~L. 2000, in Proc. SPIE
  Vol. 4013, p. 532-544, UV, Optical, and IR Space Telescopes and Instruments,
  James B. Breckinridge; Peter Jakobsen; Eds., 532--544

\bibitem[{{Lanz}(1986)}]{Lanz86}
{Lanz}, T. 1986, A\&AS, 65, 195

\bibitem[{Ma\'{\i}z~Apell\'aniz(2004)}]{Maiz04c}
Ma\'{\i}z~Apell\'aniz, J. 2004, PASP, 116, 859

\bibitem[{Ma\'{\i}z~Apell\'aniz(2005{\natexlab{a}})}]{Maiz05b}
---. 2005{\natexlab{a}}, PASP, 117, 615

\bibitem[{Ma\'{\i}z~Apell\'aniz(2005{\natexlab{b}})}]{Maiz05c}
Ma\'{\i}z~Apell\'aniz, J. 2005{\natexlab{b}}, in Resolved Stellar Populations,
  D. Valls-Gabaud and M. Ch\'avez (eds.), San Francisco: ASP (astro-ph/0506278)

\bibitem[{{Matsushima}(1969)}]{Mats69}
{Matsushima}, S. 1969, ApJ, 158, 1137

\bibitem[{{M\'egessier}(1995)}]{Mege95}
{M\'egessier}, C. 1995, A\&A, 296, 771

\bibitem[{{Mermilliod} {et~al.}(1997){Mermilliod}, {Mermilliod}, \&
  {Hauck}}]{Mermetal97}
{Mermilliod}, J.-C., {Mermilliod}, M., \& {Hauck}, B. 1997, A\&AS, 124, 349

\bibitem[{{Mignard}(2005)}]{Mign05}
{Mignard}, F. 2005, in ESA SP-576: The Three-Dimensional Universe with Gaia,
  5--14

\bibitem[{Proffitt(2005)}]{Prof05}
Proffitt, C. 2005, in 2005 HST Calibration Workshop

\bibitem[{{Skrutskie} {et~al.}(1997)}]{Skruetal97}
{Skrutskie}, M.~F., {et~al.} 1997, in ASSL Vol. 210: The Impact of Large Scale
  Near-IR Sky Surveys, F. Garz\'on et al. (eds.) (Dordrecht: Kluwer Academic
  Publishers), 25

\bibitem[{{Smith} {et~al.}(2002)}]{Smitetal02}
{Smith}, J.~A., {et~al.} 2002, AJ, 123, 2121

\bibitem[{{Stetson}(2005)}]{Stet05}
{Stetson}, P.~B. 2005, PASP, 117, 563

\bibitem[{{Str{\" o}mgren}(1966)}]{Stro66}
{Str{\" o}mgren}, B. 1966, ARA\&A, 4, 433

\bibitem[{STScI(1998)}]{synphot}
STScI. 1998, Synphot User's Guide, Howard Bushouse and Bernie Simon (eds.)

\bibitem[{{York} {et~al.}(2000)}]{Yorketal00}
{York}, D.~G., {et~al.} 2000, AJ, 120, 1579

\end{thebibliography}

\appendix

\section{Photon- and energy-counting detectors}

	When computing synthetic magnitudes from spectrophotometry, one has to be careful with an aspect that has
generated some confusion in the past. For a {\em photon-counting} detector, such as a CCD, given the
total-system dimensionless sensitivity function $P(\lambda)$, the SED of the object $f_\lambda(\lambda)$, the SED 
of the reference spectrum $f_{\lambda{\rm,ref}}(\lambda)$, and the zero point ZP$_P$ for filter $P$, the 
corresponding magnitude is given by:

\begin{equation}
m_{P} = -2.5\log_{10}\left(\frac{\int P(\lambda)f_{\lambda}(\lambda)\lambda\,d\lambda}
                                {\int P(\lambda)f_{\lambda{\rm,ref}}(\lambda)\lambda\,d\lambda}\right)
                                + {\rm ZP}_P.
\label{photon}
\end{equation}

	The $\lambda$ inside the integrals in Eq.~\ref{photon} is present because $f_\lambda(\lambda)$ is in
units of energy per time per unit area per wavelength and one needs to multiply by $\lambda/(hc)$ in order to
convert from energy to photons (the $hc$ disappears because it is a constant that appears on both integrals). For
an {\em energy-counting} (or energy-integrating) detector, the $\lambda$ is not present and the corresponding 
magnitude is given by:

\begin{equation}
m_{P}^\prime = -2.5\log_{10}\left(\frac{\int P^\prime(\lambda)f_{\lambda}(\lambda)\,d\lambda}
                                {\int P^\prime(\lambda)f_{\lambda{\rm,ref}}(\lambda)\,d\lambda}\right)
                                + {\rm ZP}^\prime_{P^\prime},
\label{energy}
\end{equation}

\noindent where $P^\prime(\lambda)$ and ZP$^\prime_{P^\prime}$ are the sensitivity function and the zero point,
respectively. 

	It is easy to verify that one can use either Eq.~\ref{photon} or Eq.~\ref{energy} independently of 
the physical type of detector involved if the sensitivity function is changed accordingly. 
For example, suppose I want to use
Eq.~\ref{photon} with an energy-counting detector. If one multiplies and divides by $\lambda$ the quantities inside 
the two integrals in Eq.~\ref{energy} and defines $P(\lambda) = P^\prime(\lambda)/\lambda$ and 
ZP$_{P}$ = ZP$^\prime_{P^\prime}$ to be the photon-equivalent sensitivity function and zero point, respectively,
then Eq.~\ref{photon} is recovered (this is the reason why previous works, such as \citealt{Bessetal98}, can be 
correct even though they use energy-integrating sensitivities). 

	The problem that has arisen sometimes is that Eq.~\ref{photon} has been used with an energy-counting
sensitivity function without first dividing it by $\lambda$ to convert it to a photon-equivalent sensitivity
function. The likely origin of the confusion has been the old practice of defining sensitivity functions for
photomultipliers as energy-counting combined with the present extensive use of ``black-box'' synthetic photometry
codes such as {\tt synphot} designed for photon-counting detectors.

\section{Absolute and relative zero points}

	ZP$_P$, the zero point used in the previous section is a relative zero point, since a reference spectrum 
is used to define it. In a number of magnitude systems (typically, those defined with synthetic photometry in 
mind), such as VEGAMAG, STMAG, or ABMAG \citep{synphot}, ZP$_P = 0.0$ by definition and the calibration information is
contained in the reference spectrum alone. For other magnitude systems derived from standard stars, ZP$_P$ is not 
exactly zero and has to be measured (that is one of the main purposes of this paper), although it is 
possible that if the right reference spectrum is used (e.g. Vega) ZP$_P$ can be indeed approximately zero.
However, one can also avoid altogether the reference spectrum in the definition of a magnitude system by using
(for a photon-counting detector):

\begin{equation}
m_{P} = -2.5\log_{10}\left(\int P(\lambda)f_{\lambda}(\lambda)\lambda\,d\lambda\right) + {\rm AZP}_P,
\label{photon2}
\end{equation}

\noindent where AZP$_P$ is the absolute zero point for filter $P$, which is related to ZP$_P$ by:

\begin{equation}
{\rm AZP}_P = 2.5\log_{10}\left(\int P(\lambda)f_{\lambda{\rm,ref}}(\lambda)\lambda\,d\lambda\right) + {\rm ZP}_P.
\label{absrel}
\end{equation}

	Note that the quantities inside the parenthesis in Eqs.~\ref{photon2}~and~\ref{absrel} are not
dimensionless (they have dimensions of energy per time per length), so the units used must be specified when 
values for AZP$_P$ are given.

\begin{deluxetable}{ccccc}
\tablecaption{Recommended photon-counting sensitivity curves for the Tycho \bt\vt\ system}
\tablewidth{0pt}
\tablehead{\multicolumn{2}{c}{\bt}        & & \multicolumn{2}{c}{\vt}        \\
           \cline{1-2}                        \cline{4-5}                      
           $\lambda$ (\AA) & $P(\lambda)$ & & $\lambda$ (\AA) & $P(\lambda)$  }
\startdata
   3500 &   0.000 &  &    4550 &   0.000 \\
   3550 &   0.014 &  &    4600 &   0.022 \\
   3600 &   0.058 &  &    4650 &   0.115 \\
   3650 &   0.123 &  &    4700 &   0.302 \\
   3700 &   0.206 &  &    4750 &   0.532 \\
   3750 &   0.305 &  &    4800 &   0.740 \\
   3800 &   0.416 &  &    4850 &   0.873 \\
   3850 &   0.530 &  &    4900 &   0.944 \\
   3900 &   0.636 &  &    4950 &   0.977 \\
   3950 &   0.724 &  &    5000 &   0.994 \\
   4000 &   0.787 &  &    5050 &   1.000 \\
   4050 &   0.830 &  &    5100 &   0.995 \\
   4100 &   0.861 &  &    5150 &   0.979 \\
   4150 &   0.889 &  &    5200 &   0.953 \\
   4200 &   0.920 &  &    5250 &   0.920 \\
   4250 &   0.953 &  &    5300 &   0.882 \\
   4300 &   0.982 &  &    5350 &   0.840 \\
   4350 &   1.000 &  &    5400 &   0.797 \\
   4400 &   0.976 &  &    5450 &   0.752 \\
   4450 &   0.861 &  &    5500 &   0.707 \\
   4500 &   0.685 &  &    5550 &   0.661 \\
   4550 &   0.489 &  &    5600 &   0.614 \\
   4600 &   0.317 &  &    5650 &   0.567 \\
   4650 &   0.202 &  &    5700 &   0.520 \\
   4700 &   0.136 &  &    5750 &   0.473 \\
   4750 &   0.101 &  &    5800 &   0.426 \\
   4800 &   0.080 &  &    5850 &   0.381 \\
   4850 &   0.059 &  &    5900 &   0.336 \\
   4900 &   0.036 &  &    5950 &   0.294 \\
   4950 &   0.016 &  &    6000 &   0.255 \\
   5000 &   0.003 &  &    6050 &   0.219 \\
   5050 &   0.000 &  &    6100 &   0.187 \\
\nodata & \nodata &  &    6150 &   0.160 \\
\nodata & \nodata &  &    6200 &   0.136 \\
\nodata & \nodata &  &    6250 &   0.114 \\
\nodata & \nodata &  &    6300 &   0.097 \\
\nodata & \nodata &  &    6350 &   0.082 \\
\nodata & \nodata &  &    6400 &   0.069 \\
\nodata & \nodata &  &    6450 &   0.058 \\
\nodata & \nodata &  &    6500 &   0.047 \\
\nodata & \nodata &  &    6550 &   0.038 \\
\nodata & \nodata &  &    6600 &   0.028 \\
\enddata
\label{tthroughputs}
\end{deluxetable}

\begin{deluxetable}{ccccccccccc}
\tablecaption{Recommended photon-counting sensitivity curves for the Str\"omgren $uvby$ standard system}
\tablewidth{0pt}
\tablehead{\multicolumn{2}{c}{$u$}        & & \multicolumn{2}{c}{$v$}        & & \multicolumn{2}{c}{$b$}        & & \multicolumn{2}{c}{$y$}        \\
           \cline{1-2}                        \cline{4-5}                        \cline{7-8}                        \cline{10-11}                    
           $\lambda$ (\AA) & $P(\lambda)$ & & $\lambda$ (\AA) & $P(\lambda)$ & & $\lambda$ (\AA) & $P(\lambda)$ & & $\lambda$ (\AA) & $P(\lambda)$  }
\startdata
   3150 &   0.000 &  &    3750 &   0.000 &  &    4350 &   0.000 &  &    5150 &   0.000 \\
   3175 &   0.000 &  &    3775 &   0.003 &  &    4375 &   0.009 &  &    5175 &   0.013 \\
   3200 &   0.007 &  &    3800 &   0.008 &  &    4400 &   0.021 &  &    5200 &   0.033 \\
   3225 &   0.076 &  &    3825 &   0.018 &  &    4425 &   0.035 &  &    5225 &   0.054 \\
   3250 &   0.220 &  &    3850 &   0.031 &  &    4450 &   0.051 &  &    5250 &   0.079 \\
   3275 &   0.424 &  &    3875 &   0.048 &  &    4475 &   0.079 &  &    5275 &   0.136 \\
   3300 &   0.607 &  &    3900 &   0.064 &  &    4500 &   0.109 &  &    5300 &   0.199 \\
   3325 &   0.758 &  &    3925 &   0.100 &  &    4525 &   0.173 &  &    5325 &   0.296 \\
   3350 &   0.879 &  &    3950 &   0.162 &  &    4550 &   0.267 &  &    5350 &   0.452 \\
   3375 &   0.961 &  &    3975 &   0.268 &  &    4575 &   0.428 &  &    5375 &   0.637 \\
   3400 &   0.999 &  &    4000 &   0.410 &  &    4600 &   0.644 &  &    5400 &   0.793 \\
   3425 &   1.000 &  &    4025 &   0.609 &  &    4625 &   0.861 &  &    5425 &   0.879 \\
   3450 &   0.990 &  &    4050 &   0.813 &  &    4650 &   0.979 &  &    5450 &   0.926 \\
   3475 &   0.972 &  &    4075 &   0.960 &  &    4675 &   1.000 &  &    5475 &   0.964 \\
   3500 &   0.945 &  &    4100 &   1.000 &  &    4700 &   0.959 &  &    5500 &   1.000 \\
   3525 &   0.898 &  &    4125 &   0.970 &  &    4725 &   0.815 &  &    5525 &   0.982 \\
   3550 &   0.830 &  &    4150 &   0.878 &  &    4750 &   0.597 &  &    5550 &   0.872 \\
   3575 &   0.742 &  &    4175 &   0.748 &  &    4775 &   0.375 &  &    5575 &   0.672 \\
   3600 &   0.634 &  &    4200 &   0.568 &  &    4800 &   0.236 &  &    5600 &   0.479 \\
   3625 &   0.503 &  &    4225 &   0.367 &  &    4825 &   0.147 &  &    5625 &   0.302 \\
   3650 &   0.357 &  &    4250 &   0.224 &  &    4850 &   0.084 &  &    5650 &   0.193 \\
   3675 &   0.239 &  &    4275 &   0.135 &  &    4875 &   0.053 &  &    5675 &   0.127 \\
   3700 &   0.157 &  &    4300 &   0.080 &  &    4900 &   0.032 &  &    5700 &   0.080 \\
   3725 &   0.107 &  &    4325 &   0.053 &  &    4925 &   0.025 &  &    5725 &   0.048 \\
   3750 &   0.068 &  &    4350 &   0.039 &  &    4950 &   0.019 &  &    5750 &   0.026 \\
   3775 &   0.039 &  &    4375 &   0.027 &  &    4975 &   0.015 &  &    5775 &   0.018 \\
   3800 &   0.019 &  &    4400 &   0.014 &  &    5000 &   0.008 &  &    5800 &   0.015 \\
   3825 &   0.000 &  &    4425 &   0.007 &  &    5025 &   0.004 &  &    5825 &   0.008 \\
\nodata & \nodata &  &    4450 &   0.000 &  &    5050 &   0.000 &  &    5850 &   0.000 \\
\enddata
\label{sthroughputs}
\end{deluxetable}

\begin{deluxetable}{cccccccc}
\tablecaption{Recommended photon-counting sensitivity curves for the Johnson $UBV$ standard system}
\tablewidth{0pt}
\tablehead{\multicolumn{2}{c}{$U$}        & & \multicolumn{2}{c}{$B$}        & & \multicolumn{2}{c}{$V$}        \\
           \cline{1-2}                        \cline{4-5}                        \cline{7-8}                      
           $\lambda$ (\AA) & $P(\lambda)$ & & $\lambda$ (\AA) & $P(\lambda)$ & & $\lambda$ (\AA) & $P(\lambda)$  }
\startdata
   3050 &   0.000 &  &    3600 &   0.000 &  &    4700 &   0.000 \\
   3100 &   0.237 &  &    3650 &   0.011 &  &    4750 &   0.004 \\
   3150 &   0.403 &  &    3700 &   0.033 &  &    4800 &   0.032 \\
   3200 &   0.489 &  &    3750 &   0.058 &  &    4850 &   0.084 \\
   3250 &   0.504 &  &    3800 &   0.144 &  &    4900 &   0.172 \\
   3300 &   0.508 &  &    3850 &   0.348 &  &    4950 &   0.310 \\
   3350 &   0.511 &  &    3900 &   0.601 &  &    5000 &   0.478 \\
   3400 &   0.513 &  &    3950 &   0.817 &  &    5050 &   0.650 \\
   3450 &   0.516 &  &    4000 &   0.958 &  &    5100 &   0.802 \\
   3500 &   0.528 &  &    4050 &   1.000 &  &    5150 &   0.913 \\
   3550 &   0.603 &  &    4100 &   0.995 &  &    5200 &   0.978 \\
   3600 &   0.741 &  &    4150 &   0.995 &  &    5250 &   1.000 \\
   3650 &   0.889 &  &    4200 &   0.994 &  &    5300 &   0.994 \\
   3700 &   0.985 &  &    4250 &   0.976 &  &    5350 &   0.977 \\
   3750 &   1.000 &  &    4300 &   0.950 &  &    5400 &   0.950 \\
   3800 &   0.965 &  &    4350 &   0.921 &  &    5450 &   0.911 \\
   3850 &   0.841 &  &    4400 &   0.888 &  &    5500 &   0.862 \\
   3900 &   0.648 &  &    4450 &   0.845 &  &    5550 &   0.806 \\
   3950 &   0.424 &  &    4500 &   0.792 &  &    5600 &   0.747 \\
   4000 &   0.231 &  &    4550 &   0.733 &  &    5650 &   0.690 \\
   4050 &   0.109 &  &    4600 &   0.673 &  &    5700 &   0.634 \\
   4100 &   0.035 &  &    4650 &   0.619 &  &    5750 &   0.579 \\
   4150 &   0.000 &  &    4700 &   0.569 &  &    5800 &   0.523 \\
\nodata & \nodata &  &    4750 &   0.519 &  &    5850 &   0.467 \\
\nodata & \nodata &  &    4800 &   0.467 &  &    5900 &   0.413 \\
\nodata & \nodata &  &    4850 &   0.414 &  &    5950 &   0.363 \\
\nodata & \nodata &  &    4900 &   0.362 &  &    6000 &   0.317 \\
\nodata & \nodata &  &    4950 &   0.315 &  &    6050 &   0.274 \\
\nodata & \nodata &  &    5000 &   0.272 &  &    6100 &   0.234 \\
\nodata & \nodata &  &    5050 &   0.232 &  &    6150 &   0.200 \\
\nodata & \nodata &  &    5100 &   0.193 &  &    6200 &   0.168 \\
\nodata & \nodata &  &    5150 &   0.155 &  &    6250 &   0.140 \\
\nodata & \nodata &  &    5200 &   0.121 &  &    6300 &   0.114 \\
\nodata & \nodata &  &    5250 &   0.096 &  &    6350 &   0.089 \\
\nodata & \nodata &  &    5300 &   0.075 &  &    6400 &   0.067 \\
\nodata & \nodata &  &    5350 &   0.054 &  &    6450 &   0.050 \\
\nodata & \nodata &  &    5400 &   0.034 &  &    6500 &   0.037 \\
\nodata & \nodata &  &    5450 &   0.018 &  &    6550 &   0.027 \\
\nodata & \nodata &  &    5500 &   0.007 &  &    6600 &   0.020 \\
\nodata & \nodata &  &    5550 &   0.001 &  &    6650 &   0.016 \\
\nodata & \nodata &  &    5600 &   0.000 &  &    6700 &   0.013 \\
\nodata & \nodata &  & \nodata & \nodata &  &    6750 &   0.012 \\
\nodata & \nodata &  & \nodata & \nodata &  &    6800 &   0.010 \\
\nodata & \nodata &  & \nodata & \nodata &  &    6850 &   0.009 \\
\nodata & \nodata &  & \nodata & \nodata &  &    6900 &   0.007 \\
\nodata & \nodata &  & \nodata & \nodata &  &    6950 &   0.004 \\
\nodata & \nodata &  & \nodata & \nodata &  &    7000 &   0.000 \\
\enddata
\label{jthroughputs}
\end{deluxetable}

\begin{deluxetable}{lcccccccc}
\tablecaption{Color/index zero points and associated uncertainties/errors.}
\tablewidth{0pt}
\tablehead{ & Tycho-2 & & \multicolumn{3}{c}{Str\"omgren} & & \multicolumn{2}{c}{Johnson} \\
              \cline{2-2} \cline{4-6}                         \cline{8-9} 
            & \btvt   & & \by   & \m1   & \c1             & & \bv   & \ub                 }
\startdata
zero point  & 0.033   & & 0.007 & 0.154 & 1.092           & & 0.010 & 0.020               \\
random      & 0.001   & & 0.001 & 0.001 & 0.002           & & 0.001 & 0.006               \\
systematic  & 0.005   & & 0.003 & 0.003 & 0.004           & & 0.004 & 0.014               \\
\enddata
\label{colorzp}
\end{deluxetable}

\begin{deluxetable}{lccccccrrr}
\tablecaption{Magnitude zero points and associated uncertainties. The values for $y$ are from \citet{HolbBerg06} and those
for the 2MASS filters from \citet{Coheetal03}.}
\tablewidth{0pt}
\tablehead{  & Tycho-2       & & Str\"omgren   & & Johnson       & & \multicolumn{3}{c}{2MASS}                            \\
               \cline{2-2}       \cline{4-4}       \cline{6-6}       \cline{8-10} 
             & \colhead{\vt} & & \colhead{$y$} & & \colhead{$V$} & & \colhead{$J$} & \colhead{$H$} & \colhead{$K_{\rm s}$}}
\startdata
zero point   & 0.034         & & 0.014         & & 0.026         & & $-$0.001      & 0.019         & $-$0.017             \\
uncertainty  & 0.006         & & 0.010         & & 0.008         & &    0.005      & 0.007         &    0.005             \\
\enddata
\label{magnitudezp}
\end{deluxetable}

\begin{deluxetable}{lccc}
\tablecaption{Dispersions for $(\by)_{\rm phot} - (\by)_{\rm spec}$ as a function of 
$V$ and $N$ for all the data points (excluding 3.5 sigma outliers).}
\tablewidth{0pt}
\tablehead{ & \multicolumn{3}{c}{$N$}         \\
              \cline{2-4}                       
            & $\le 3$ & $\ge 4$ & All}
\startdata
         $V \le 6.0$ & 0.012 & 0.011 & 0.012 \\
           $V > 6.0$ & 0.014 & 0.013 & 0.014 \\
                 All & 0.014 & 0.013 & 0.013 \\
\enddata
\label{bysigmas}
\end{deluxetable}

\begin{deluxetable}{lccc}
\tablecaption{Dispersions for $\m1_{, \rm phot} - \m1_{, \rm spec}$ as a function of 
$V$ and $N$ for all the data points (excluding 3.5 sigma outliers).}
\tablewidth{0pt}
\tablehead{ & \multicolumn{3}{c}{$N$}         \\
              \cline{2-4}                       
            & $\le 3$ & $\ge 4$ & All}
\startdata
         $V \le 6.0$ & 0.013 & 0.012 & 0.012 \\
           $V > 6.0$ & 0.015 & 0.013 & 0.014 \\
                 All & 0.014 & 0.013 & 0.014 \\
\enddata
\label{m1sigmas}
\end{deluxetable}

\begin{deluxetable}{lccc}
\tablecaption{Dispersions for $\c1_{, \rm phot} - \c1_{, \rm spec}$ as a function of 
$V$ and $N$ for all the data points (excluding 3.5 sigma outliers).}
\tablewidth{0pt}
\tablehead{ & \multicolumn{3}{c}{$N$}         \\
              \cline{2-4}                       
            & $\le 3$ & $\ge 4$ & All}
\startdata
         $V \le 6.0$ & 0.018 & 0.015 & 0.017 \\
           $V > 6.0$ & 0.019 & 0.018 & 0.018 \\
                 All & 0.018 & 0.017 & 0.018 \\
\enddata
\label{c1sigmas}
\end{deluxetable}

\begin{deluxetable}{lccc}
\tablecaption{Dispersions for $(\bv)_{\rm phot} - (\bv)_{\rm spec}$ as a function of 
$V$ and $N$ for all the data points (excluding 3.5 sigma outliers).}
\tablewidth{0pt}
\tablehead{ & \multicolumn{3}{c}{$N$}         \\
              \cline{2-4}                       
            & $\le 3$ & $\ge 4$ & All}
\startdata
         $V \le 6.0$ & 0.017 & 0.018 & 0.018 \\
           $V > 6.0$ & 0.020 & 0.024 & 0.022 \\
                 All & 0.019 & 0.020 & 0.020 \\
\enddata
\label{bvsigmas}
\end{deluxetable}

\begin{deluxetable}{lcccc}
\tablecaption{Dispersions for $(\ub)_{, \rm phot} - (\ub)_{, \rm spec}$ as a function of 
$V$ and $N$ for all the data points (excluding 3.5 sigma outliers).}
\tablewidth{0pt}
\tablehead{ & \multicolumn{3}{c}{$N$}         \\
              \cline{2-4}                       
            & $\le 3$ & $\ge 4$ & All}
\startdata
         $V \le 6.0$ & 0.026 & 0.029 & 0.028 \\
           $V > 6.0$ & 0.030 & 0.027 & 0.029 \\
                 All & 0.028 & 0.028 & 0.028 \\
\enddata
\label{ubsigmas}
\end{deluxetable}

\end{document}